\documentclass[transmag]{IEEEtran}
\usepackage{latexsym}
\usepackage{graphicx}
\usepackage{amsfonts,amssymb,amsmath}
\usepackage{hyperref}

\usepackage{amsthm}
\usepackage{url}

\usepackage{bbm}
\usepackage{multirow}

\usepackage{array}

\usepackage{suffix}
\usepackage{mathtools}

\usepackage{float}
\usepackage{color}
\usepackage{cite}

\ifCLASSOPTIONcompsoc
\usepackage[caption=false,font=normalsize,labelfont=sf,textfont=sf]{subfig}
\else
\usepackage[caption=false,font=footnotesize]{subfig}
\fi

\DeclarePairedDelimiterX\MeijerM[3]{\lparen}{\rparen}%
{\begin{smallmatrix}#1 \\ #2\end{smallmatrix}\delimsize\vert\,#3}

\newcommand\MeijerG[8][]{%
	G^{\,#2,#3}_{#4,#5}\MeijerM[#1]{#6}{#7}{#8}}

\WithSuffix\newcommand\MeijerG*[7]{%
	G^{\,#1,#2}_{#3,#4}\MeijerM*{#5}{#6}{#7}}

\DeclareMathOperator\erfc{erfc}  

\newcounter{eqcnt}

\def\BibTeX{{\rm B\kern-.05em{\sc i\kern-.025em b}\kern-.08em T\kern-.1667em\lower.7ex\hbox{E}\kern-.125emX}}
\begin{document}

\title{Statistical Modeling of the Impact of Underwater Bubbles on an Optical Wireless Channel}
\author{Myoungkeun Shin,~\IEEEmembership{Nonmember},
	Ki-Hong Park,~\IEEEmembership{Senior Member,~IEEE,} and 
	Mohamed-Slim Alouini,~\IEEEmembership{Fellow,~IEEE}
	\thanks{M. Shin, K.-H. Park, and M.-S. Alouini are with the Computer, Electrical,
		and Mathematical Science and Engineering (CEMSE) Division, King Abdullah University of Science and Technology (KAUST), Thuwal, Makkah
		Province, 23955-6900 Kingdom of Saudi Arabia (e-mail: {myoungkeun.shin; kihong.park;
			slim.alouini}@kaust.edu.sa).}
}

\IEEEtitleabstractindextext{\begin{abstract}
In underwater wireless optical communications (UWOC), the random obstruction of light propagation by air bubbles can cause fluctuations in the incoming light intensity of a receiver. In this paper, we propose a statistical model for determining the received power by a receiver in the presence of air bubbles. First, based on real experiments of the behavior of air bubbles underwater, we propose statistical models for the generation, size, and horizontal distribution of each air bubble. Second, we mathematically derive the obstruction caused by the shadow of each bubble as it passes over the beam area. We then compute the combined obstruction of all generated air bubbles to determine the total obstructed power, which is a random variable due to the randomness of bubble behavior. Next, we find the first and second moments of the total obstructed power to model the statistical distribution of the obstructed received power by using the method of moments, 
which shows that the Weibull distribution suitably matches the simulation data. We also estimate the shape and scale parameters by using two derived moments.
Furthermore, we also construct a statistical model of the received power with complete blockage in the presence of air bubbles and we derive the distribution of the composite channel model combining the proposed bubble-obstruction model with a Gamma-Gamma turbulence model. Finally, we obtain and verify the analytic forms of the average bit error rate and the capacity of UWOC systems under this newly proposed composite channel model.	
\end{abstract}

\begin{IEEEkeywords}
Underwater wireless optical communications, air bubbles, turbulence, and performance analysis.
\end{IEEEkeywords}

}

\maketitle

\section{Introduction}

\IEEEPARstart{W}{ireless} communications are a crucial component of several underwater applications such as tactical surveillance, pollution monitoring, oil control and maintenance, offshore oil explorations, climate change monitoring, and oceanography research\cite{key6}.
Wireless underwater communications can be classified according to their means of communication, i.e., acoustic, radio frequency (RF), and optical.\footnote{Please see some table in more details which compares the pros and cons of acoustic and optical communications in \cite{key6}.} Conventional acoustic communications can be transmitted over long distances, but the data transfer rate is limited to several kbps/Mbps. RF waves suffer from severe attenuation during propagation. In contrast, even though underwater wireless optical communication (UWOC) can travel over short distances up to 100m only, UWOC can transfer information at much higher data rates, in the order of a Gbps~\cite{Shen16}. 

UWOC is restricted to short transmission distance because of environmental attenuation factors such as absorption and scattering. 
In this context, laser beam propagation can be characterized mathematically in the form of the radiative transfer equation (RTE)\cite{key8}, \cite{key2}, \cite{key3ex} and probabilistically by the Monte-Carlo simulation \cite{key9}. Extensive analytical, numerical and experimental research has been conducted\cite{key10}. However, although turbulence caused by random variations in temperature and salinity induces the fluctuations in beam power detected at the receiver, there has been relatively minimal research in this area. Most studies on underwater turbulence still apply the existing free-space optical (FSO) channel models such as lognormal distribution for the atmospheric environment, even though the atmospheric environment is quite different from the underwater environment. 

The presence of air bubbles represents one of the main differences between the atmospheric and the underwater environment. For example, research oceanographer Grant B. Deane at the University of California, San Diego, calls our attention to bubbles in oceans. In large bodies of water, air bubbles are generated when waves break on the shore and sea surface~\cite{bubbles}. The presence of bubbles may impact power fluctuations at the receiver. A recent study developed a statistical model of underwater turbulence induced by air bubbles based on real experiments\cite{key11,key23}. In those studies, the mixed exponential-Gamma distribution and exponential-lognormal distributions were proposed to model the statistics of received power, which is different from traditional lognormal distribution in the FSO turbulence model. Experimental results in \cite{key11} also showed that different bubble sizes and beam sizes affect power fluctuation at the receiver. Unfortunately, the probability that air bubbles perfectly block the optical propagation towards the receiver aperture was not considered in that proposed model, even though the experimental results showed that this situation can occur~\cite{key12,key23}.

To address this issue, here we newly develop a statistical model for the fluctuation of received power in UWOC systems under the randomness of air bubble generation, size and horizontal distribution. Based on these statistics of air bubbles, we calculate the expectation and the second moment of the obstructed received power. Using the method of moments, we propose a mixture of Weibull distribution and two Dirac delta functions in which the former represents partial blockage and the latter complete and no blockage. The main contributions of this work are as follows;
	\begin{itemize}
		\item Unlike the statistical model developed by experimental measurements in \cite{key11,key23}, our proposed statistical model is purely mathematical and numerically solved by using the statistics of the generation, size, and the horizontal movement of a bubble.
		\item Our proposed statistical distribution more precisely models the effect of air bubbles on beam propagation by taking into consideration the probabilities of complete blockage and no blockage which are the most striking features differentiating from the aforementioned models in \cite{key11,key23}. 
		\item We provide the closed-form statistics of received signal-to-noise ratio (SNR) over composite channel model, which combines the proposed bubble-induced fading model and a Gamma-Gamma turbulence model.
		\item Finally, we analyze the ergodic capacity and the average bit error rate (BER).
	\end{itemize}

The remainder of this paper is organized as follows. In Section II, we present the setup of the statistical model induced by air bubbles in underwater optical wireless channels. In Section III, we first delineate the overall strategy to find the fitted distribution and we then show the precision of predicted models compared with simulation data by using the mean-square-error (\textit{MSE}) and the $R^2$ tests. In Section IV, we present the results of the performance analysis of the composite model and verify its accuracy by comparison with simulation results. Finally, some concluding remarks are given in Section V.

\section{System Setup for the Statistical Modeling}

\begin{figure}[t]
	\centering
	\includegraphics[width=0.9\linewidth]{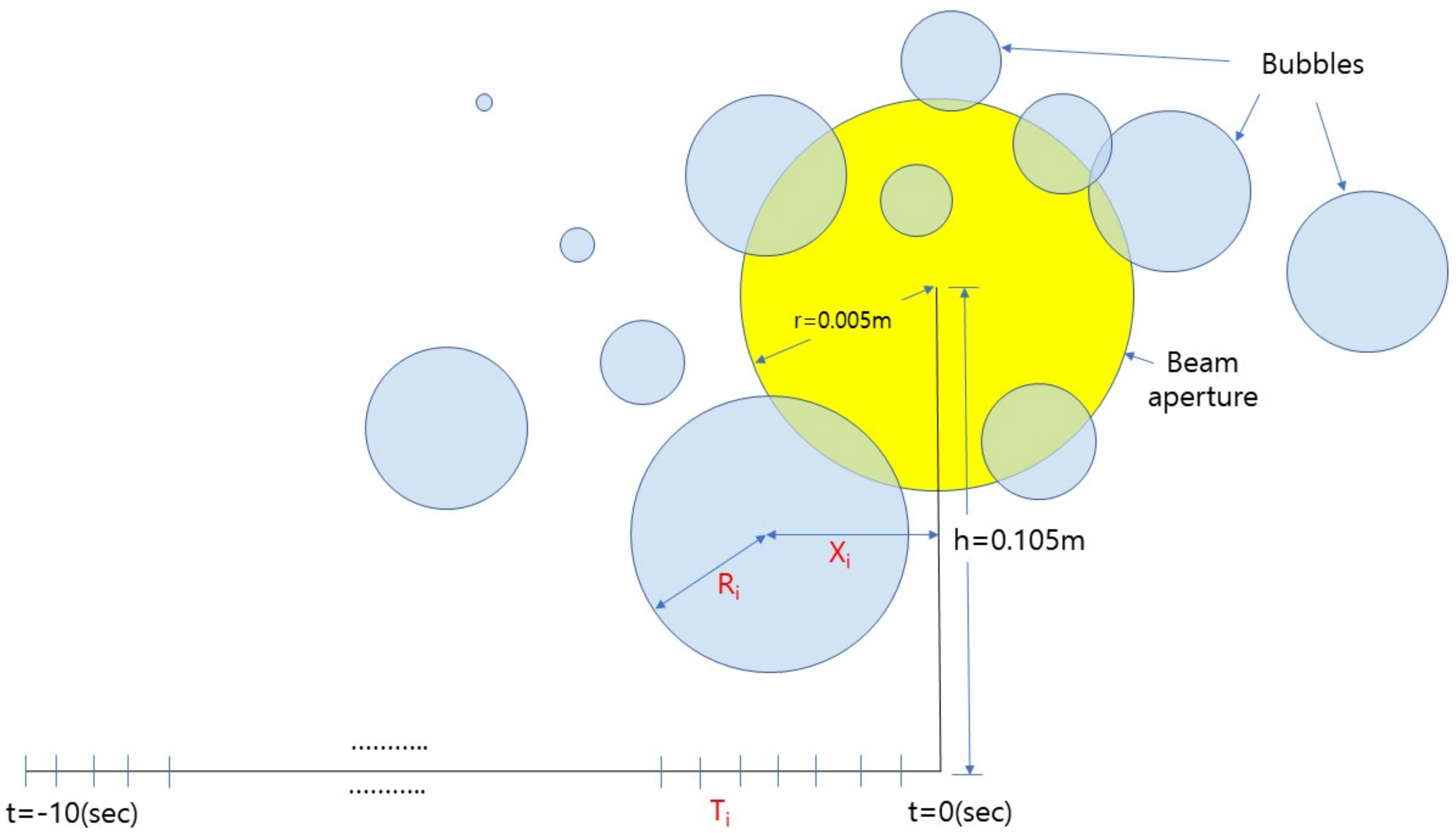}
	\caption{Example of air bubbles in a single layer rising around an optical beam on $(w,z)$-coordinate.}
	\label{fig:bubble}
\end{figure}

The system setup for our statistical model is based on the models based on two real experiments discussed in previous works \cite{key11}, \cite{key12}. We assume the system consists of a water tank with air bubbles emerging from the holes at the bottom of the tank. An optical beam is  perfectly aligned to the receiver aperture and propagated from one wall of the tank to the opposite wall, where the receiver is located.  Specifically, we assume that the optical beam has the following Gaussian distribution:
\begin{equation}
h(w,z) = \frac{1}{2\pi \sigma^2}  \exp \Big( - \frac{w^2 + z^2}{2\sigma^2}\Big),
\end{equation}
where $\sigma = 5$ [mm] is the variance related to the beam waist and $(w,z)$-coordinate spans the space perpendicular to the beam propagation. In~\cite{key11,key12}, the diameters of optical beam and active area at the photodiode at the receiver are nearly 4 [mm]. The optical beam diameter less than 10 [mm] in~\cite[Fig. 7]{key12} is shown to be reasonable to investigate the fluctuation of received power. Therefore, we assume that the beam aperture at the receiver is circular, and its radius is $r=5$ [mm]. Here, we assume that transmit laser source and receiver aperture are embedded on the stand with a height of $10$ [cm]. Accordingly, the beam center is assumed to be located at a height of 0.105 [m] from the bottom of the tank. We note that this specific value of beam center does not affect the simulation results since the time duration of generating air bubbles, which is 10 [sec], is long enough to span all the air bubbles affecting the received power regardless of a height of beam center. 
Fig.~\ref{fig:bubble} illustrates the beam propagation in the presence of air bubbles around the receiver at $(w,z)$-coordinate. We outline the assumptions of the generation rate, the horizontal movement, and the sizes of bubbles in the following subsections in more detail. 

\vspace{3ex}


\subsection{Random Generation of Air Bubbles}

We note that we calculate the received power at one time instant (current). Before that time instant, we assume that each bubble is generated uniformly within its own specific time interval, which is related to the blow rate of air bubbles in the experiment. Every interval has the same length, only one bubble is generated in each interval, and the intervals do not overlap with each other. We also assume that the power is only obstructed by the bubbles that have been generated between the current time instant and 10 [sec] before that.\footnote{The reason why we only consider 10 [sec] is that the bubbles generated before 10 [sec] do not cause any significant obstruction. More specifically, relatively big bubbles travel up the aperture faster due to their faster rising velocity, while the influence on the obstruction by relatively small bubbles is minimal.} In this sense, the probability density function (PDF) of the time instant when generating the bubble in the $i$th time interval closest to the current time instant, $T_i$ is given by 
\begin{equation}
f_{T_i}(t) = \frac{1}{L}, \quad (i-1)L \le t \le iL,
\end{equation}
where $L$ is the length of an interval. We use four different lengths for simulation: $1/20$, $1/40$, $1/80$, and $1/160$ [sec], which provide different levels of obstruction of the beam propagation. The lower $L$ is, the more bubbles are generated for the given 10 [sec] period.

\subsection{Horizontal Movement of Air Bubbles}

For simplicity, we assume only a single layer of bubbles originate from a single hole at the bottom of the tank, and the layer is parallel to the wall where the receiver is located. Although rising bubbles vibrate a bubble jitters a little when moving upwards, their horizontal movement cannot be considered as Brownian motion. It is also possible to assume multiple layers of bubbles by adding multiple independent bubbles in each interval, but in this paper, we only consider a single layer of bubbles. Thus, we assume that bubbles only move upwards and that the bubbles are distributed in a single layer according to a specific Gaussian distribution. Specifically, as shown in Fig.~\ref{fig:bubble}, the minimum distance of a bubble from the line that goes through the center of the beam and is perpendicular to the bottom of the tank is assumed to be a Gaussian random variable $X_i$ with zero mean and variance $\sigma_x^2$. We assume that the standard deviation $\sigma_x=\sigma$ is the same as the radius of the beam aperture, which is $5$ [mm]. As the standard deviation of $X_i$, $5$ [mm] is less than the actual deviations in the real experiments shown in \cite[Fig. 5]{key12}. We note that this can emphasize the effect of the obstruction due to the multiple air bubbles on light propagation. The distribution of horizontal distance $X_i$ is given by
\begin{equation}
f_{X_i}(x) = \frac{1}{\sqrt{2\pi\sigma^2}}e^{-\frac{x^2}{2\sigma^2}}.
\end{equation}

\subsection{Vertical Movement of Air Bubbles} 

Here, we describe the vertical movement of air bubbles to quantify the vertical distance of a bubble from the center of the optical beam, which is closely related to the generation time (explained in Section II.B) and the rising velocity of an air bubble, which depends on the size-dependent bubble shape. Moreover, using the upper limit on the area of a bubble shown in \cite[Fig. 5]{key12}, we assume there the bubble size does not exceed 0.01 [m]. Accordingly, the distribution of $R_i$ is represented by

\begin{equation}
	f_{R_i}(r; \mu_{R_i}) = \frac{\pi r }{2{\mu_{R_{i}}}^2}e^{\frac{-\pi r^2}{4{\mu_{R_{i}}}^2}} \Big/ q , \quad 0 \le r \le 0.01, 
\end{equation}
where $q = \int_0^{0.01} \frac{\pi r }{2{\mu_{R_{i}}}^2}e^{\frac{-\pi r^2}{4{\mu_{R_{i}}}^2}}dr$ is the normalization factor and $\mu_{R_{i}}$ is the mean of the radius of bubbles. Here, we consider four different values of $\mu_{R_i}$ for simulation, i.e., 1.36 [mm], 1.50 [mm], 1.95 [mm], and 2.99 [mm], respectively. We note that these four specific values are derived from the mean area of air bubbles in \cite[Fig. 5]{key12}.

\begin{figure}[t]
	\centering
	\includegraphics[width=0.9\linewidth]{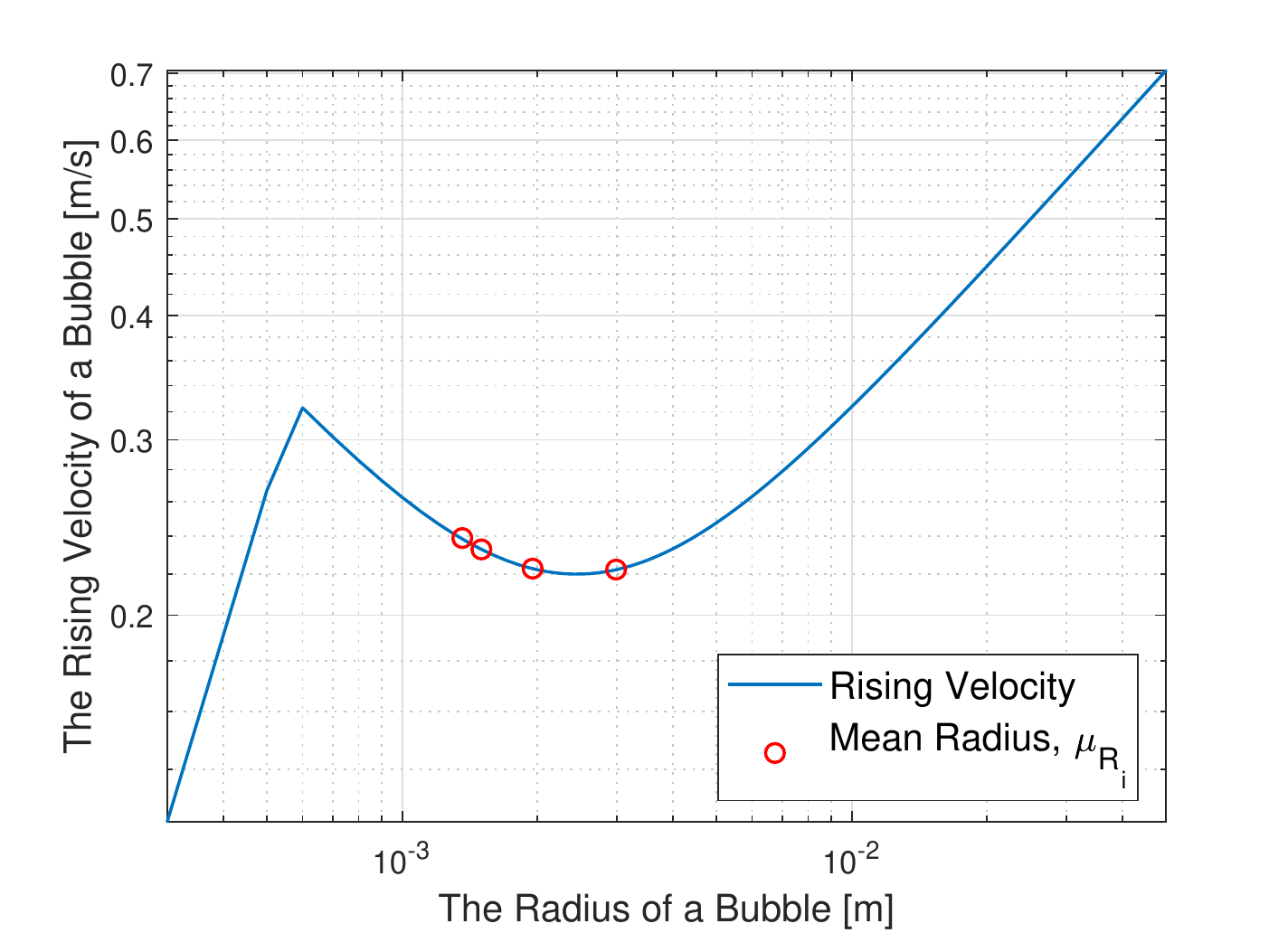}
	\caption{Rising velocity of a bubble in water~\cite{key13}.}
	\label{radius}
	\vspace{-0.2in}
\end{figure}

The rising velocity of an air bubble changes mainly because of its diameter and shape. More specifically, the bubble has three types of shape; sphere, spheroid, and spherical cap\cite{key13}. Fig.\ref{radius} shows how the rising velocity of a bubble changes as its diameter increases. When a bubble is very small, it is spherical because the inertial force is small compared to the viscous force, and its rising velocity increases as it grows in size because of the dominance of buoyancy. As its size further increases, the inertial force becomes tangible and the increasing rate of the rising velocity begins to slow down. When it becomes even bigger in size, the bubble changes its shape from sphere to spheroid, and its rising velocity starts to decrease because the friction becomes tangible. However, the rising velocity increases again as the ratio of the longer axis length to the shorter axis length increases. When the bubble becomes much bigger, the bubble changes its shape from spheroid to spherical cap. The rising velocity of the bubble, $v$ [m/s], is given by\cite{key13}

\begin{equation}
v=
\begin{cases}
\displaystyle\frac{g\rho {R_i}^2}{3\mu}, & 0 < R_i < 0.08015\textrm{ [mm]}\\
\displaystyle 0.408g^{\frac{5}{6}}\left(\frac{\rho}{\mu}\right)^{\frac{2}{3}}{R_i}^{\frac{3}{2}}, \,& 0.08015 \le  R_i  < 0.575 \textrm{ [mm]}\\
\displaystyle \sqrt{\frac{1.07\rho}{\sigma_s {R_i}}+1.01g{R_i}}, \, & R_i \ge 0.575 \textrm{ [mm]},
\end{cases}
\end{equation}
where $R_i$ [m] is the radius of a bubble, $\rho$ is the density of the liquid medium, $\sigma_s$ is the surface tension of the liquid medium, and $g$ is gravitational acceleration. In order to calculate the area in aperture obstructed by air bubbles at the specific vertical position of current time instance, we can determine the vertical position of air bubble by multiplying the rising time duration with the rising velocity related to the generated bubble's radius.
Here we remark that, even though we admit that it is more realistic to use all possible bubble shapes, it is highly difficult to calculate the overlapped areas with spheroid/spherical cap-shaped bubbles. Therefore, we assume that all bubbles are spherical when it comes to the calculation, while we use (5) as the rising velocity.

\section{Proposed Statistical Model for Bubble Obstruction}

It is very difficult to find the explicit PDF of the obstructed power due to the air bubbles directly from the statistical models of bubble behavior. Therefore, we will use the method of moments, which compares the moments of the derived statistical model and a well-known distribution, and estimates the parameters of the distribution so that their moments match each other\cite{key22}. We will go through the following procedures. 
\begin{itemize}
	\item[1)] Let the random variables $B_i$ be the amount of obstructed power due to the $i$th bubble and $B_i$ can be expressed as a function of $X_i$, $R_i$, and $T_i$.
	\begin{align}
	B_i &= b_i(X_i, R_i, T_i) \nonumber \\
	&= B_i^{(1)}  \mathbbm{1}_{\left( 0, |r-R_i|\right)}(D_i)  \mathbbm{1}_{( 0, r)}(R_i) \nonumber\\
	&  +  B_i^{(2)}  \mathbbm{1}_{\left( |r-R_i|, \sqrt{r^2-R_i^2}\right)}(D_i)  \mathbbm{1}_{( 0, r)}(R_i) \nonumber \\
	& + B_i^{(3)}  \mathbbm{1}_{\left( \sqrt{r^2-R_i^2}, (r+R_i)\right)}(D_i)  \mathbbm{1}_{( 0, r)}(R_i) \nonumber 
\end{align}
\begin{align}
	& + B_i^{(4)}  \mathbbm{1}_{\left( 0, |r-R_i|\right)}(D_i)  \mathbbm{1}_{( 0, R_i)}(r) \nonumber \\
	& +  B_i^{(5)} \mathbbm{1}_{\left( |r-R_i|, \sqrt{R_i^2-r^2}\right)}(D_i)  \mathbbm{1}_{( 0, R_i)}(r) \nonumber \\
	& + B_i^{(6)}  \mathbbm{1}_{\left( \sqrt{R_i^2-r^2}, (r+R_i)\right)}(D_i)  \mathbbm{1}_{( 0, R_i)}(r),
	\end{align}
	where $B_i^{(j)}$ $(j=1,2,\cdots, 6)$ indicates the obstructed power for the various cases of a bubble's location with respect to the beam and $D_i$ denotes the distance between the centers of a bubble and the aperture. The details of deriving each $B_i^{(j)}$ are delineated in Appendix B. Also, $\mathbbm{1}_{\mathcal{S}}(x)$ is an indicator function that is defined as 
	\begin{equation}
	\mathbbm{1}_{\mathcal{S}}(x)=
	\begin{cases}
	1, \quad x \in \mathcal{S}\\
	0, \quad x \notin \mathcal{S}.\\
	
	\end{cases}
	\end{equation}
	
	\item[2)] Defining the sum of $B_i$, $\forall i$, as a random variable $B$, we calculate the expectation and the second moment of $B$. The formulas are described in detail in Appendix C.
	\item[3)] We can model the PDF of $B$ as given by
	\begin{eqnarray}
	f_B(x)= a\delta(x)+ b f_W(x), \;x \ge 0,
	\end{eqnarray}
	where $ \delta(x)$ is the Dirac delta function, $a$ is the probability of no power obstruction, and and $b=1-a$ is the probability of any power obstruction. $f_W(x)$ is the normalized density distribution of power obstruction for $x>0$, which is modeled as the Weibull distribution given by
	\begin{equation}
	f_W(x) = 	   		        \frac{k}{\lambda}\left(\frac{x}{\lambda}\right)^{k-1} e^{(x/ \lambda)^k}, \quad x \ge 0, 
	\end{equation}
	where $\lambda$ and $k$ are the scale and shape parameters.    With the first and second moments of $B$, we estimate the scale and shape parameters of the Weibull distribution. The details are described in Appendix D. 
	\item[4)] The power obstructed by air bubbles cannot exceed the maximum received power onto the beam aperture, i.e., $m = \int_\mathcal{A} h(w,z)\,dw\,dz$, where $\mathcal{A}$ is the beam aperture area without any obstruction. Then, we modify the PDF of $B$ to the following form 
	\begin{align}
	f_B(x)= a\delta(x)+ bf_W(x) + c\delta(x-m), \;0 \le x \le m,
	\end{align}
	where $c = \int_m^{\infty} b f_W(x)dx$ indicates the complete obstruction of the received power.
	\item[5)] Finally, we can obtain the distribution of the received power in the presence of air bubble obstruction at the receiver as 
	\begin{equation}
	f_{H_b}(x) = c\delta(x)+ bf_W(m - x) + a\delta(x-m), \;0 \le x \le m.\label{pdf_hb}
	\end{equation}

\end{itemize}

\subsection{The Results of the Parameter Estimation in $f_{H_b}(x)$}

In Table~\ref{weiresult}, we provide the values of parameters $a$, $b$ and $c$ when we use the Gaussian beam that we described in the previous sections. We also provide the comparison of the estimated distribution $f_{h_b}(x)$ and the simulation data. For the simulation data, $a$ means the probability of no power obstruction and $c$ means the complete obstruction. 

\begin{table*}[t!]
	\renewcommand{\arraystretch}{1.3}
	\caption{The Parameters and Test Results for Bubble Generation Rates of 20, 40, 80 and 160 [1/sec]}
	\begin{center}
		\label{tab:table1}
		\begin{tabular}{|c|c|c|c|c|c|c|c|c|c|c|c|}
			\hline 
			\multicolumn{2}{|c|}{\textbf{Distribution}} & \multicolumn{3}{c|}{\textbf{Simulation Data}} & \multicolumn{7}{c|}{\textbf{Proposed Distribution}}  \\
			\hline
			\multirow{2}{*}{\textbf{Rate} \textrm{[1/sec]}} & \multirow{2}{*}{\textbf{Radius} \textrm{[mm]}} & \multicolumn{3}{c|}{\textbf{Parameters}} & \multicolumn{5}{c|}{\textbf{Parameters}} &
			\multicolumn{2}{c|}{\textbf{Test Results}} \\
			\cline{3-12}
\multirow{2}{*}{} & \multirow{2}{*}{} & $c$ & $b$ & $a$ & $c$ & $b$ & $a$ & $k$ & $\lambda$ & {\textbf{MSE}} & $R^2$ \\
			\hline
			\multirow{4}{*}{\textbf{20}} & 1.35 & 0 & 0.58 & 0.42 & {3.28E-6} & 0.58 & 0.42 & 0.935 & 0.030 & 3.43E-5 & 0.997 \\
			\multirow{4}{*}{} & 1.50 & 5.00E-5 & 0.60 & 0.40 & 1.41E-5 & 0.60 & 0.40 & 0.955 & 0.036 & 3.27E-5 & 0.998 \\ 
			\multirow{4}{*}{} & 1.95 & 6.30E-4 & 0.67 & 0.33 & 4.92E-4 & 0.66 & 0.34 & 0.968 & 0.054 & 2.50E-5 & 0.998 \\
			\multirow{4}{*}{} & 2.99 & 0.02 & 0.74 & 0.26 & 0.01 & 0.75 & 0.25 & 0.983 & 0.102 & 1.30E-4 & 0.996 \\
			\hline
			\multirow{4}{*}{\textbf{40}} & 1.35 & 0 & 0.82 & 0.18 & 2.36E-5 & 0.82 & 0.18 & 1.307 & 0.044 & 6.97E-5 & 0.985 \\
			\multirow{4}{*}{} & 1.50 & 1.40E-4 & 0.84 & 0.16 & 1.04E-4 & 0.84 & 0.16 & 1.064 & 0.053 & 6.61E-5 & 0.987 \\
			\multirow{4}{*}{} & 1.95 & 3.40E-3 & 0.89 & 0.11 & 2.80E-3 & 0.89 & 0.11 & 1.110 & 0.085 & 3.88E-5 & 0.990 \\
			\multirow{4}{*}{} & 2.99 & 0.07 & 0.94 & 0.06 & 0.06 & 0.94 & 0.06 & 1.170 & 0.172 & 1.92E-4 & 0.986 \\
			\hline
			\multirow{4}{*}{\textbf{80}} & 1.35 & 3.90E-4 & 0.97 & 0.03 & 2.16E-4  & 0.97 & 0.03 & 1.290 & 0.079 & 5.98E-5 & 0.944 \\
			\multirow{4}{*}{} & 1.50 & 1.50E-3 & 0.97 & 0.03 & 1.00E-3 & 0.98 & 0.02 & 1.400 & 0.097 & 5.18E-5 & 0.941 \\
			\multirow{4}{*}{} & 1.95 & 0.03 & 0.99 & 0.01 & 0.03 & 0.99 & 0.01 & 1.448 & 0.162 & 1.79E-5 & 0.976 \\
			\multirow{4}{*}{} & 2.99 & 0.29 & 1 & 3.80E-3 & 0.28 & 1 & 3.30E-3 & 1.590 & 0.343 & 1.67E-5 & 1.000 \\
			\hline
			\multirow{4}{*}{\textbf{160}} & 1.35 & 0.01 & 1 & 7.50E-4 & 4.70E-3 & 1 & 9.25E-4 & 1.806 & 0.160 & 6.38E-6 & 0.971 \\
			\multirow{4}{*}{} & 1.50 & 0.03 & 1 & 5.70E-4 & 0.02 & 1 & 5.96E-4 & 1.892 & 0.197 & 1.18E-5 & 0.978 \\
			\multirow{4}{*}{} & 1.95 & 0.22 & 1 & 1.50E-4 & 0.22 & 1 & 1.27E-4 & 2.087 & 0.328 & 9.0E-5 & 0.998 \\
			\multirow{4}{*}{} & 2.99 & 0.77 & 1 & 0 & 0.76 & 1 & 1.03E-5 & 0.231 & 0.692 & 4.41E-5 & 1.000 \\
			\hline 
		\end{tabular}
	\end{center}
	\label{weiresult}
\end{table*}

\begin{figure*}[t!]
	\centering
	\subfloat[Radius($\mu_R$): 1.35 {[mm]}]{\includegraphics[width=0.25\textwidth]{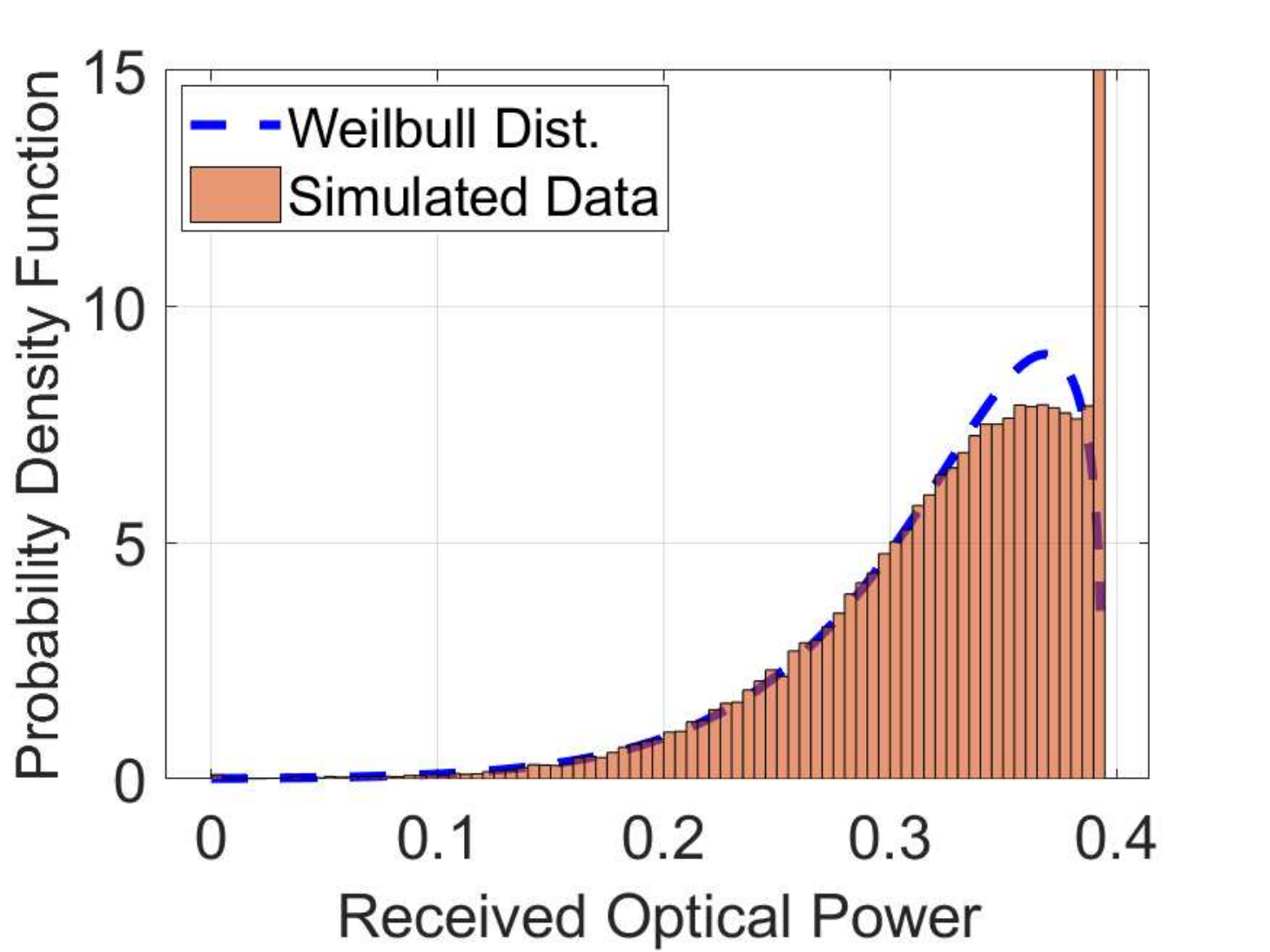}
		\label{radius1_rate3}}
	\subfloat[Radius($\mu_R$): 1.50 {[mm]}]{\includegraphics[width=0.25\textwidth]{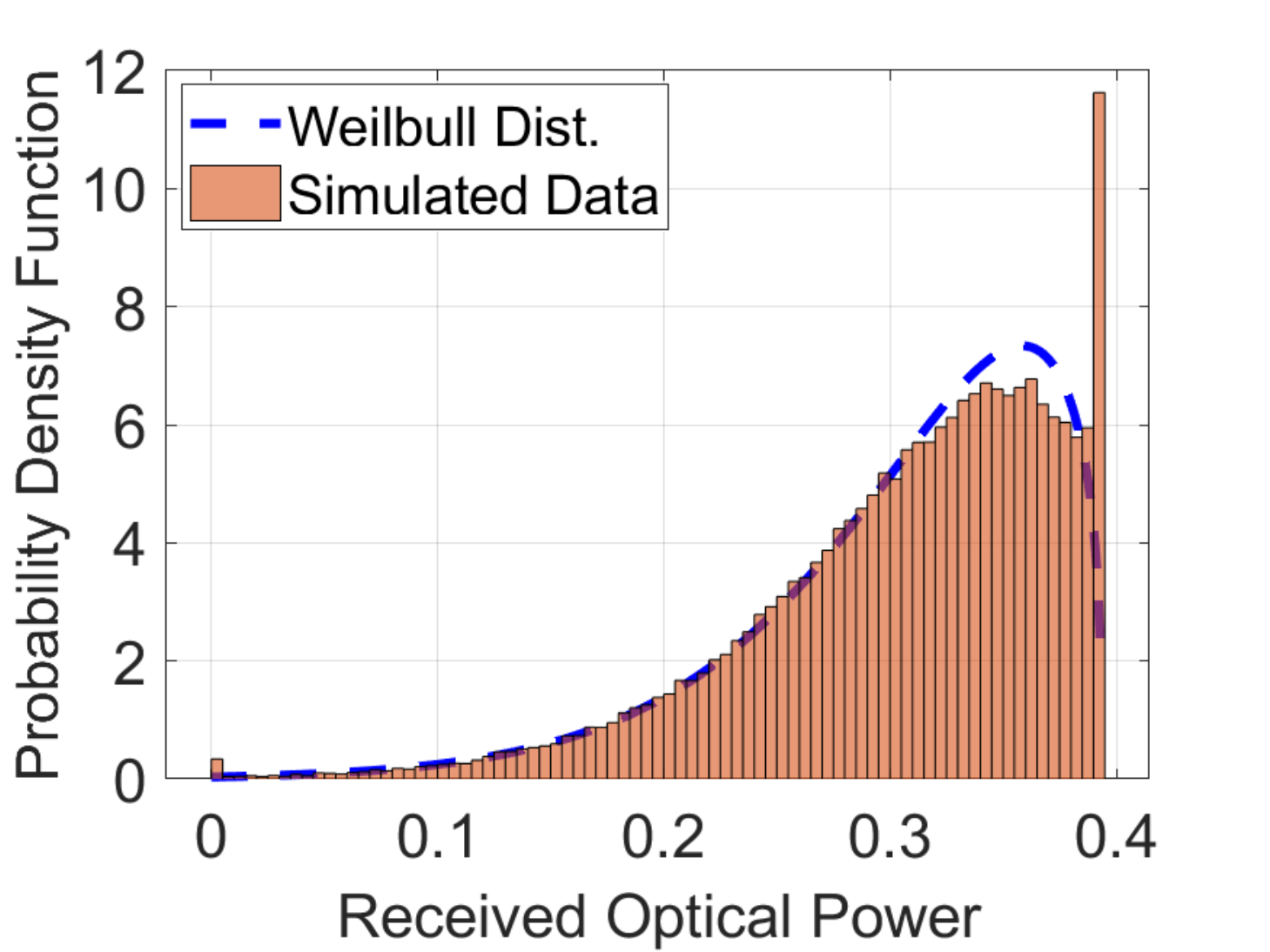}
		\label{radius2_rate3}}
	\subfloat[Raduis($\mu_R$): 1.95 {[mm]}]{\includegraphics[width=0.25\textwidth]{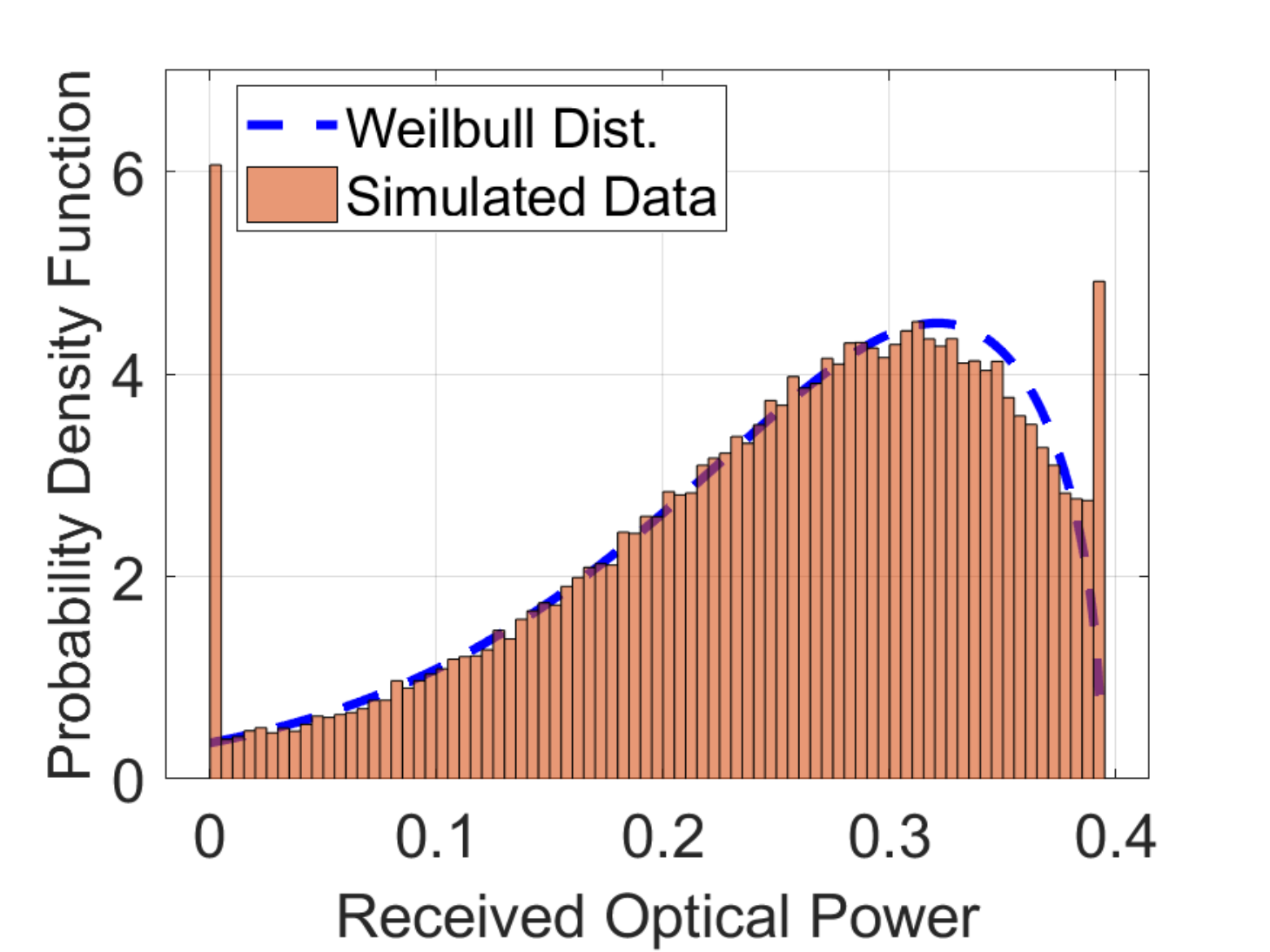}
		\label{radius3_rate3}}
	\subfloat[Radius($\mu_R$): 2.99 {[mm]}]{\includegraphics[width=0.25\textwidth]{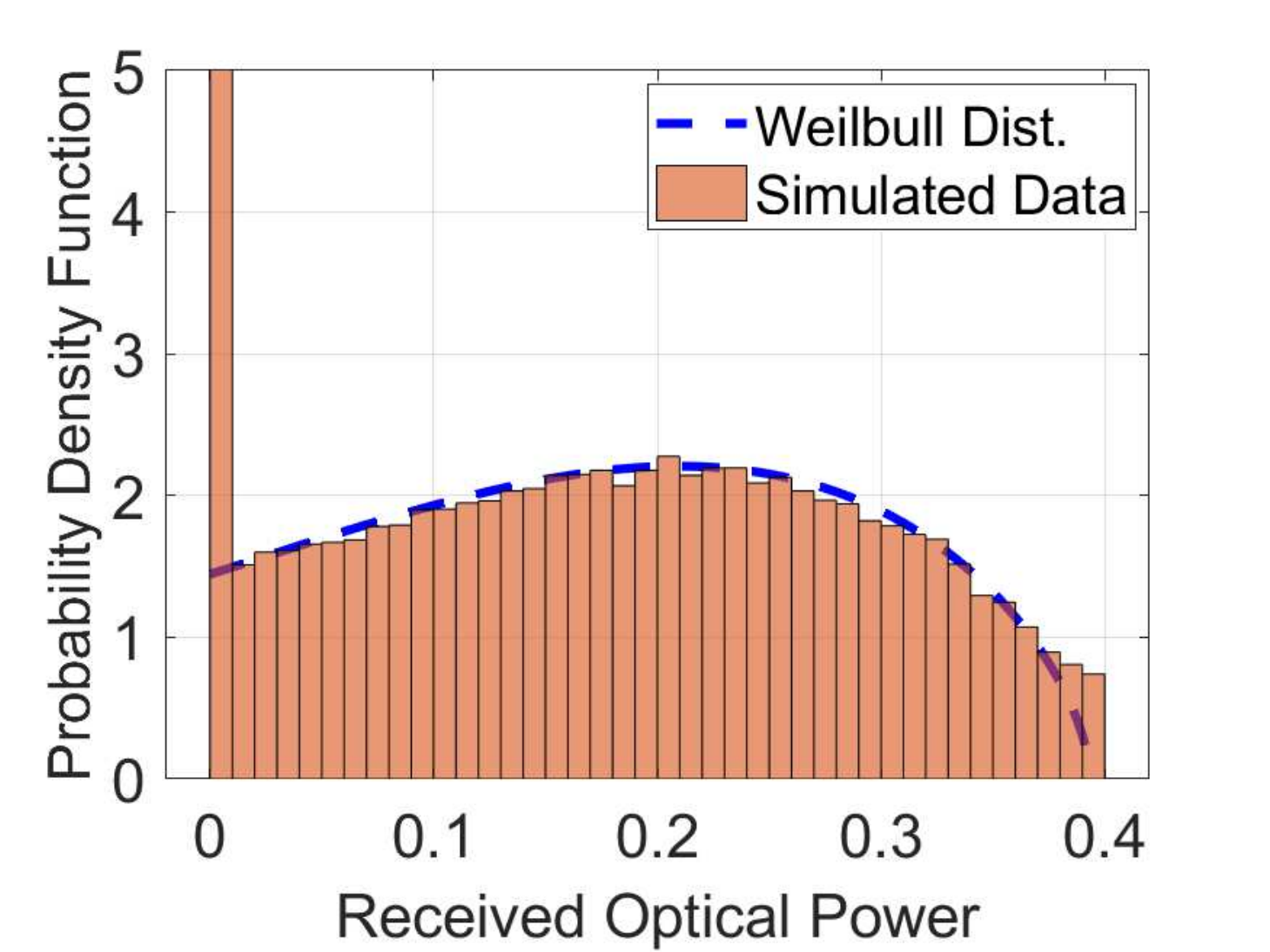}
		\label{radius4_rate3}}
	\caption{The distributions of the four different radii of bubbles when the bubble generation rate is 80 [1/sec]. The areas of first and last peak bars in histogram of simulated data indicate the probabilities of complete blockage ($c$) and no blockage ($a$) which is not drawn for our proposed statistical model. }
	\label{rate3}
\end{figure*}

In all four cases of the bubble generation rate, as the average radius of the bubbles increases, the chance of power obstruction increases. Furthermore, as the generation rate increases with the average increase in the radius, there is increased chance of power obstruction. When the average of the radius is $1.35$ [mm], and the rate is $20$ [1/sec], the probability of no power obstruction is above $0.42$. On the other hand, when the average of the radius is $2.99$ [mm], and the rate is $160$ [1/sec], the probability of complete power failure is $0.76$. We can check that the values of the parameters $a$, $b$, and $c$ are very close to  simulation data. From the comparison between our proposed statistical model and simulation data, we can see the relation between the system parameters considered in our model and the estimated parameters for blockage on the beam propagation as in Table~\ref{acresult} below.

\begin{table}[H]
		\renewcommand{\arraystretch}{1.3}
		\caption{The tendency of estimated parameters with respect to the system parameters}
		\begin{center}
			\label{tab:table_ac}
			\begin{tabular}{|c|c|c|}
				\hline 
				\multirow{2}{*}{\textbf{System Parameters}}  & \multicolumn{2}{c|}{\textbf{Estimated Parameters}}\\
				\cline{2-3}
				\multirow{2}{*}{} & Perfect blockage $c$  & No blockage $a$    \\
				\hline
				Radius of Aperture $r$ ($\nearrow$) & $\searrow$ & $\searrow$   \\
				\hline
				Radius of Air Bubble $\mu_{R_i}$ ($\nearrow$) & $\nearrow$ & $\searrow$   \\
				\hline
				Horizontal Movement $\sigma_x$ ($\nearrow$) & $\searrow$ & $\nearrow$   \\
				\hline
				Bubble Arrival $L$ ($\nearrow$) & $\searrow$ & $\nearrow$  \\
				\hline 
			\end{tabular}
		\end{center}
		\label{acresult}
\end{table} 

\subsection {The Results of Fit Tests}

In this section, we present the goodness of fit for two tests, which are the $MSE$ test and the $R^2$ test. The definition of the $MSE$ test is given by 

\begin{equation}
MSE = \frac{\sum^N_{i=1}[F_s(I_i) - F(I_i)]^2}{N},
\end{equation}
where $F_s(I_i)$ is the value of the accumulated probability function of the simulation data at certain points $I_i$ and $F(I_i)$ is the value of the cumulative distribution function (CDF) of the distribution, $f_{h_b}(x)$, at certain points $I_i$. When it comes to the goodness of fit of the \textit{MSE} test, the closer the value is to $0$, the better the fit is. 

The $R^2$ test metric is defined as
\begin{equation}
R^2 = 1 - \frac{S_e}{S_t},
\end{equation}
where $S_e = \sum_{i=1}^M (f_{s,i} - f_{p_i})^2$ and $S_t = \sum_{i=1}^M (f_{s,i} - \bar{f})^2$, $f_{s,i}$ and $f_{p,i}$ are the probability of the simulated data and $f_{h_b}(x)$ at the $i$th interval, $\bar{f}$ is the mean of $f_{s,i}$ and $M$ is the number of the intervals. When it comes to the goodness of fit of the $R^2$ test, the closer the value is to $1$, the better the fit is.

Table \ref{weiresult} shows the results of goodness of fit tests for our proposed model. For all cases, the \textit{MSE} test results are close to 0 and the $R^2$ test results are close to 1. We note that the distribution of the received power is the lower bound of actual received power because we calculate the sum of obstructed powers between every bubble and the beam area but we neglect the fact that the bubbles cannot overlap. The discrepancy between actual distribution and the proposed model increases as the generation rate and average size of air bubbles increase.

\subsection{Comparison between Simulation and Statistical Model}

Fig. \ref{rate3} shows how the simulation data and its corresponding distribution are drawn, specifically for the cases of four different bubble radii with a bubble generation speed of $80$ [1/sec]. We note in Fig. \ref{rate3} that the areas of first and last peak bars in histogram of simulated data converge to the probabilities of complete blockage ($c$) and no blockage ($a$) as the bin size decreases. For our proposed statistical model, we only draw the continuous part regarding Weibull distribution.

We can see that the simulation data matches its predicted distribution. The left tail of each distribution in particular matches very well, which is very important to predict average BER. Moreover, we note that the cases with no obstruction and full obstruction were considered, which were ignored in the previous work of \cite{key11}.

\section{Performance Analysis}

Underwater turbulence due to salinity and temperature variations can commonly cause a fluctuation in the received power. To evaluate the performance in UWOC channels in the presence of air bubbles and underwater turbulence, we combine the distribution of our proposed bubble obstruction model with another independent turbulence distribution. We first derive the composite PDF of received signal-to-noise ratio (SNR) over air bubble obstruction and Gamma-Gamma turbulence and analyze the ergodic capacity and average BER. The received signal can be written as
\begin{equation}
y = hx + n,
\end{equation}
where $x$ is the transmitted intensity, $h$ is the channel state, and $n$ is additive white Gaussian noise (AWGN) with zero mean and variance $\sigma_n^2$. The channel state $h$ consists of three factors, i.e., $h=h_lh_ah_b$ with path loss $h_l$, turbulence $h_a$, and bubble obstruction $h_b$. These factors are independent of each other, and $h_l$ is deterministic, whereas $h_a$ and $h_b$ follow the aforementioned statistical distributions. Here we assume that the beam is a Gaussian beam perfectly aligned to the receiver aperture and the turbulence fading distribution is Gamma-Gamma fading channel for strong turbulence underwater environments. We now derive the statistics of received SNR over composite channel models by combining our proposed bubble obstruction model with the Gamma-Gamma fading model.

\subsection{Statistics of Received SNR}

We can write the instantaneous received SNR as follows:
\begin{equation}
\gamma = \frac{\vert h \vert^2  \mathbb{E}[x^2]}{\mathbb{E}[n^2]} = \vert h \vert^2  \bar{\gamma},
\end{equation}
where $\bar{\gamma} = \frac{\mathbb{E}[x^2]}{\mathbb{E}[n^2]}$ is defined as average SNR. Since $\bar{\gamma}$ and $h_l$ are constant, we first find the CDF of $h_{ab}=h_ah_b$, which is computed as
\begin{eqnarray}
F_{H_{ab}}(x) \!\!&\!\!=\!\!&\!\! \Pr(H_a H_b \le x) \nonumber\\
\!\!&\!\!=\!\!&\!\! \int^{\infty}_{0}\!\!\!F_{H_a}\!\!\left(\frac{x}{y}\right)\!\!  f_{H_b}(y)  dy \nonumber \\
\!\!&\!\!=\!\!&\!\! \int^{\infty}_{0}\!\!\!F_{H_a}\!\!\left(\frac{x}{y}\right)\!\!  \left(c \delta(y)\! + \!b  f_W(1 \!-\! y) \!+\! a  \delta (y\!-\!1) \right) dy \nonumber \\
\!\!&\!\!=\!\!&\!\! c \!+\! b \int^{1}_{0} \!\!\!F_{H_a}\!\!\left(\frac{x}{y}\right) \! f_W(1\! - \!y) \, d y + a F_{H_a}(x),
\end{eqnarray}
where we note that $m = 1$ due to the average SNR $\bar{\gamma}$, and $F_{H_a}(\cdot)$ is the CDF of a Gamma-Gamma distribution whose PDF is given by $f_{H_a}(x) = \frac{2(\alpha \beta)^{(\alpha + \beta)/2}}{\Gamma(\alpha) \Gamma(\beta)}x^{\frac{\alpha + \beta}{2}-1} K_{\alpha -\beta}\left(2\sqrt{\alpha \beta x}\right)$ where $K_\alpha(x)$ is the modified Bessel function of the second kind \cite{key18}.

Due to the analytical intractability of integration, we use the Gauss-Legendre quadrature method to approximate the second term as follows:
\begin{align}
 &\int^{1}_{0} F_{H_a}\left(\frac{x}{y}\right)  f_W(1 - y) dy \nonumber \\
 & \approx  \frac{1}{2}\sum^{n}_{i=1} w_i F_{H_a}\left(\frac{2x}{x_i+1}\right)f_W\left(\frac{1-x_i}{2}\right),
\end{align}
where $w_i$ and $x_i$ are the weight and the abscissa, and $w_i =\frac{2}{(1-x_i^2)[P_n^{'}(x_i)]^2}$ with Legendre polynomials $P_n(x)$. Therefore, 
\begin{equation}
F_{H_{ab}}(x) \approx c + \frac{b}{2} \sum^{n}_{i=1} w_i F_{H_a}\!\!\left(\frac{2x}{x_i\!+\!1}\right)\!f_W\!\left(\frac{1\!-\!x_i}{2}\right) + a F_{H_a}(x).
\end{equation}

Accordingly, the channel gain $h=h_lh_al_b$ and the received SNR $\gamma$ can be computed by change of variables as 
\begin{equation}
F_{H}(x) \approx  c + \frac{b}{2} \! \sum^{n}_{i=1}\! w_i F_{H_a}\!\!\left(\!\frac{2x}{h_l(x_i\!+\!1)}\!\right)\!f_W\!\!\left(\!\frac{1\!-\!x_i}{2}\!\right) + a  F_{H_a}\!\!\left(\!\frac{x}{h_l}\!\right) 
\end{equation}
and
\begin{align}
F_\gamma(x) &\approx  c + \frac{b}{2} \sum^{n}_{i=1} w_i F_{H_a}\left(\frac{2\sqrt{x}}{h_l\sqrt{\bar{\gamma}}(x_i+1)}\right)f_W\left(\frac{1-x_i}{2}\right) \nonumber \\
& + a  F_{H_a}\left(\frac{\sqrt{x}}{h_l\sqrt{\bar{\gamma}}}\right), 
\end{align}
respectively. Therefore, the PDF of $\gamma$ is given by 
\begin{align}
f_\gamma(x) &\approx \; c \, \delta(x) + \frac{b}{2} \sum^{n}_{i=1} C^{(0)}_i x^{-1/2} f_{H_a}\left(\frac{2\sqrt{x}}{h_l\sqrt{\bar{\gamma}}(x_i+1)}\right) \nonumber \\ 
& + a C^{(0)}_0 x^{-1/2} f_{H_a}\left(\frac{\sqrt{x}}{h_l\sqrt{\bar{\gamma}}}\right),
\end{align}
where $C^{(0)}_i = \frac{w_i}{h_l\sqrt{\bar{\gamma}}(x_i+1)} f_W\left(\frac{1-x_i}{2}\right)$ and $C^{(0)}_0 = \frac{1}{2 h_l\sqrt{\bar{\gamma}}}$.


\subsection{Ergodic Capacity}

The ergodic channel capacity $\bar{C}$ is defined as $\bar{C} = \mathbb{E}[\log_2(1+\gamma)]$. We represent $\log(1+\gamma)$ as $
\MeijerG*{1}{2}{2}{2}{1,1}{1,0}{\gamma}$ and the modified Bessel function of the second kind $K_\nu(x)$ as $\frac{1}{2}\MeijerG*{2}{0}{0}{2}{-}{\frac{\nu}{2},-\frac{\nu}{2}}{\frac{x^2}{4}}$  using Meijer's G function. \cite{key19}
\begin{align}
\mathbb{E}[\log_2(1+\gamma)] =& \,\int^{\infty}_{0} \log_2(1+\gamma) f_\gamma(\gamma) \, d\gamma \nonumber \\
\approx &  \sum^{n}_{i=1} C^{(1)}_i \int^{\infty}_{0} \gamma^{\frac{\alpha + \beta}{4}-1} \MeijerG*{1}{2}{2}{2}{1,1}{1,0}{\gamma} \nonumber \\
\times & \MeijerG*{2}{0}{0}{2}{-}{\frac{\alpha - \beta}{2},-\frac{\alpha - \beta}{2}}{\frac{2\alpha\beta}{h_l\sqrt{\bar{\gamma}}(x_i+1)} \gamma^{1/2}}\, d\gamma \nonumber \\
+& \, C^{(1)}_0\int^{\infty}_{0} \gamma^{(\frac{\alpha + \beta}{4}-1)} \MeijerG*{1}{2}{2}{2}{1,1}{1,0}{\gamma} \nonumber \\
\times & \MeijerG*{2}{0}{0}{2}{-}{\frac{\alpha - \beta}{2},-\frac{\alpha - \beta}{2}}{\frac{\alpha\beta}{h_l\sqrt{\bar{\gamma}}} \gamma^{1/2}}\, d\gamma,
\end{align}
where $C^{(1)}_i = \frac{bC^{(0)}_i}{2\log2} \frac{(\alpha \beta)^{(\alpha + \beta)/2}}{\Gamma(\alpha) \Gamma(\beta)}  \left(\frac{2}{h_l\sqrt{\bar{\gamma}}(x_i+1)} \right)^{\frac{\alpha + \beta}{2} -1 } $ and $C^{(1)}_0 = \frac{aC^{(0)}_0}{\log2} \frac{(\alpha \beta)^{(\alpha + \beta)/2}}{\Gamma(\alpha) \Gamma(\beta)}  \left(\frac{1}{h_l\sqrt{\bar{\gamma}}} \right)^{\frac{\alpha + \beta}{2} -1 }$. Using \cite[Eq. (21)]{key21}, we can express the ergodic capacity with Meijer's G function, which is given on the top of next page by \eqref{EC},
\begin{figure*}
	\begin{align}
	\bar{C} \approx &  \sum^{n}_{i=1}\frac{C^{(1)}_i}{2\pi} \MeijerG*{6}{1}{2}{6}{-\frac{\alpha + \beta}{4},1 -\frac{\alpha + \beta}{4}}{\frac{\alpha - \beta}{4},\frac{\alpha - \beta +2}{4},-\frac{\alpha - \beta}{4},-\frac{\alpha - \beta -2}{4},-\frac{\alpha + \beta}{4},-\frac{\alpha + \beta}{4} }{\frac{{z_i}^2}{16}} \nonumber \\
	+& \, \frac{C^{(1)}_0}{2\pi} \MeijerG*{6}{1}{2}{6}{-\frac{\alpha + \beta}{4},1 -\frac{\alpha + \beta}{4}}{\frac{\alpha - \beta}{4},\frac{\alpha - \beta +2}{4},-\frac{\alpha - \beta}{4},-\frac{\alpha - \beta -2}{4},-\frac{\alpha + \beta}{4},-\frac{\alpha + \beta}{4}}{\frac{{z_0}^2}{16}}. \label{EC}
	\end{align}
	\hrule
\end{figure*}
where $z_i = {\frac{2\alpha\beta}{h_l\sqrt{\bar{\gamma}}(x_i+1)}}$ and $z_0 ={\frac{\alpha\beta}{h_l\sqrt{\bar{\gamma}}}} $.

\subsection{Average BER}

The average BER is defined as $P_b =\mathbb{E}[ Q(p \sqrt{q\gamma})]$, where $Q(\cdot)$ is the Q-function and $p$ and $q$ are modulation-depending parameters. Here, we use $p=1$ and $q=2$ for simulation. We note that $Q(x) =  \frac{1}{2}\erfc\left(\frac{x}{\sqrt{2}}\right)$ and $\erfc(\sqrt{x})$ can be expressed as $\erfc(\sqrt{x}) = \frac{1}{\sqrt{\pi}}\MeijerG*{2}{0}{1}{2}{1}{0,\frac{1}{2}}{x}$ in the form of Meijer's G function.\cite{key20}  Then, we can derive the average BER as
\begin{align}
P_b =&\;\; \int^{\infty}_{0} Q(p \sqrt{q\gamma}) f_\gamma(\gamma) \, d\gamma \nonumber\\
=& \;\;\frac{1}{2}\int^{\infty}_{0}  \erfc\left(p \sqrt{\frac{q\gamma}{2}}\right)f_\gamma(\gamma) \, d\gamma \nonumber \\
\approx &\;\; \frac{c}{2} + \sum^{n}_{i=1} C^{(2)}_i \int^{\infty}_{0} \gamma^{\frac{\alpha + \beta}{4}-1} \MeijerG*{2}{0}{1}{2}{1}{0,\frac{1}{2}}{\frac{p^2 q}{2}\gamma} \nonumber \\
\times & \MeijerG*{2}{0}{0}{2}{-}{\frac{\alpha - \beta}{2},-\frac{\alpha - \beta}{2}}{\frac{2\alpha\beta}{h_l\sqrt{\bar{\gamma}}(x_i+1)} \gamma^{1/2}}\, d\gamma \nonumber \\
+&\;\; C^{(2)}_0 \int^{\infty}_{0} \gamma^{\frac{\alpha + \beta}{4}-1} \MeijerG*{2}{0}{1}{2}{1}{0,\frac{1}{2}}{\frac{p^2 q}{2}\gamma} \nonumber \\
\times & \MeijerG*{2}{0}{0}{2}{-}{\frac{\alpha - \beta}{2},-\frac{\alpha - \beta}{2}}{\frac{\alpha\beta}{h_l\sqrt{\bar{\gamma}}} \gamma^{1/2}}\, d\gamma, \label{BER}
\end{align} 
where $C^{(2)}_i = \frac{bC^{(0)}_i}{2\sqrt{\pi}} \frac{(\alpha \beta)^{(\alpha + \beta)/2}}{\Gamma(\alpha) \Gamma(\beta)}    \left(\frac{2}{h_l\sqrt{\bar{\gamma}}(x_i+1)} \right)^{\frac{\alpha + \beta}{2} -1 } $ and $C^{(2)}_0 = \frac{a C^{(0)}_0}{\sqrt{\pi}} \frac{(\alpha \beta)^{(\alpha + \beta)/2}}{\Gamma(\alpha) \Gamma(\beta)}  \left(\frac{1}{h_l\sqrt{\bar{\gamma}}} \right)^{\frac{\alpha + \beta}{2} -1 } $. Using \cite[Eq. 21]{key21}, we can express the average BER with Meijer's G function on the top of next page in \eqref{BER1}.

\begin{figure*}
\begin{align}
P_b \approx &\;\; \frac{c}{2} + \sum^{n}_{i=1}\frac{C^{(2)}_i}{4\pi} \left({\frac{p^2 q}{2}}\right)^{-\frac{\alpha + \beta}{4}} \MeijerG*{4}{2}{2}{5}{1-\frac{\alpha + \beta}{4},\frac{1}{2} -\frac{\alpha + \beta}{4}}{\frac{\alpha - \beta}{4},\frac{\alpha - \beta +2}{4},-\frac{\alpha - \beta}{4},-\frac{\alpha - \beta -2}{4},-\frac{\alpha + \beta}{4} }{\frac{{z_i}^2}{8p^2 q}} \nonumber\\
+&\;\; \frac{C^{(2)}_i}{4\pi}  \left({\frac{p^2 q}{2}}\right)^{-\frac{\alpha + \beta}{4}} \MeijerG*{4}{2}{2}{5}{1-\frac{\alpha + \beta}{4},\frac{1}{2} -\frac{\alpha + \beta}{4}}{\frac{\alpha - \beta}{4},\frac{\alpha - \beta +2}{4},-\frac{\alpha - \beta}{4},-\frac{\alpha - \beta -2}{4},-\frac{\alpha + \beta}{4} }{\frac{{z_0}^2}{8p^2 q}}.
\label{BER1}
\end{align}
\hrule
\end{figure*}

\subsection{Numerical Results}

In this subsection, we validate our analytic results with comparison to the simulation results. For simulation set-up, we assume that the path loss for signal attenuation is assumed to be normal, i.e., $h_l=1$ without loss of generality and we use $\alpha = 2.21$, $\beta = 3.31$ for the two parameters of the Gamma-Gamma distribution from \cite{key15}.\footnote{We note that
another set of parameters $(\alpha, \beta)$ for different underwater turbulence related to the salinity and temperature variation can be found by using the methods in \cite{key16,key17}.} The simulated results are evaluated with random samples of composite channel gains generated by multiplying our simulation data for air bubble obstruction with Gamma-Gamma distributed random realizations. We compare the performance results obtained by using this PDF with those using the
corresponding simulation data, which are shown in Figs. \ref{performance}-\ref{performance_dev}. Throughout the figures, we note that the analytic results derived with our proposed statistical model fit well with the simulation data.

Fig.~\ref{performance} presents the results of performance analysis as a function of mean radius of air bubble. In both sub-figures, we see that the complete obstruction due to the air bubbles can significantly degrade both performance metrics. In Fig.~\ref{performance}(a), we observe that the perfect blockage with the probability value $c$ will affect the slope of ergodic capacity with respect to SNR, also known as multiplexing gain or degree of freedom. With the results of average BER in~\ref{performance}(b), we see that each BER has an error floor with a value $\frac{c}{2}$ drawn in black color. In other words, we can see the average BER in (\ref{BER1}) converges with $\frac{c}{2}$ as SNR tends to infinity. This implies that when the air bubbles exist in the path of the UWOC link, they cause a fade in SNR over time until they start to rise. In that case, the design the compensation techniques to mitigate the effect of deep fading by air bubbles is required, e.g., forward error correction (FEC), automatic repeat request (ARQ), interlearver, etc. 

Fig.~\ref{performance_rate} illustrates the results of both performance metrics with respect to different bubble generation rates. As the bubble generation rate increases, the ergodic capacity and average BER become worse. This is because more bubbles might probably occur on beam propagation path due to the increased generation of air bubbles. On other front, we show in Fig.~\ref{performance_dev} the effect of system performance for different value of standard deviation for horizontal movement of air bubbles. We see that increased horizontal movement can help improving both performances. When the rising bubbles are more scattered due to the variation of water flow and pressure, the bubbles might be spread out over beam aperture in probability and thus the obstruction can be reduced.

\section{Conclusion}

In this paper, we developed a statistical model describing the received power obstructed by air bubbles in a UWOC channel. Our proposed model is based on three random variables: which are the horizontal movement, the size and the generation of a bubble. Based on the statistics of those random behavior of air bubbles, we derived the received power and its moments. By using the method of moments, we find suitable distribution fitting the simulation data. We derived  the distribution of the received power as a combination of the well-known Weibull distribution and two of the Dirac delta functions. We verified goodness of fit for our proposed statistical model using the \textit{MSE} and the $R^2$ tests. Finally, we combined our statistical model of air bubble obstruction with a Gamma-Gamma turbulence model and analyzed the performance over this composite channel model. We confirmed that the approximated analytic results were well matched with the simulation results.


\appendix[Statistical Model Using Method of Moments]
\renewcommand{\theequation}{A.\arabic{equation}}
\setcounter{eqcnt}{\value{equation}}
\setcounter{equation}{0}

\subsection{Notations}

\begin{itemize}
	\item[(1)] The radius of bubble: $R_i$ 
	\item[(2)] The radius of the receiver aperture: $r$
	\item[(3)] The distance between the center of a bubble and the line that is perpendicular to the bottom of the tank and goes through the center of the beam: $X_i$
	\item[(4)] The time duration until the current time instant of a bubble generated in the $i$ time interval in the past: $T_i$
	\item[(5)] The height of a bubble: $H_i = h_i(T_i,R_i)=v(R_i)T_i$
	\item[(6)] The distance between the center of a bubble and that of the beam: $D_i = d_i(X_i,H_i)=\sqrt{X_i^2+(H_i-0.105)^2}$
	\item[(7)] The distribution of the Gaussian beam: $h(w,z)$
	\item[(8)] The power obstructed by each bubble: $B_i = b_i(X_i, R_i, T_i)$
\end{itemize}

\clearpage
\newpage

\begin{figure*}[h]
	\centering
	\subfloat[Ergodic Capacity]{\includegraphics[width=3.5in]{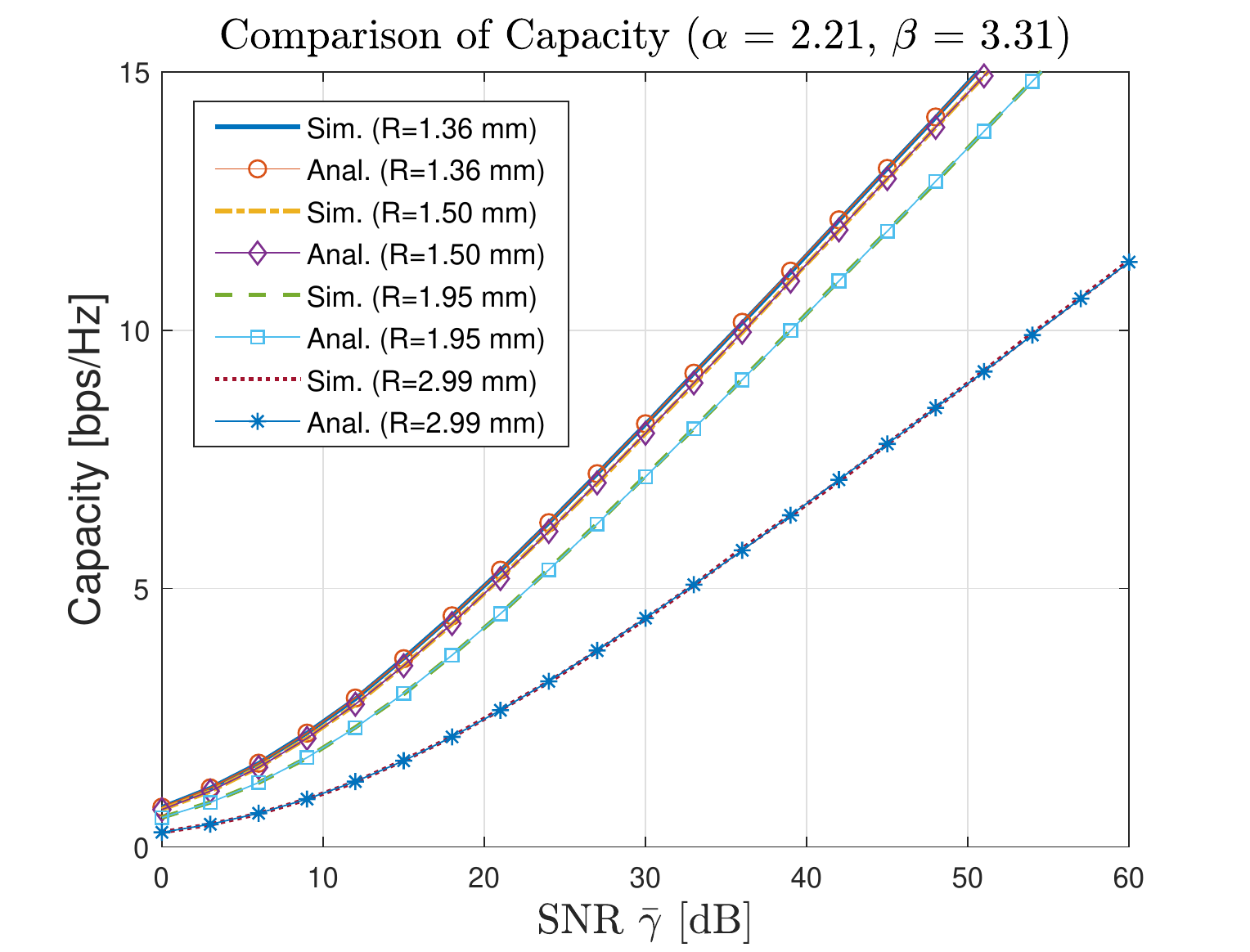}
	}
	\hfil
	\subfloat[Average BER]{\includegraphics[width=3.5in]{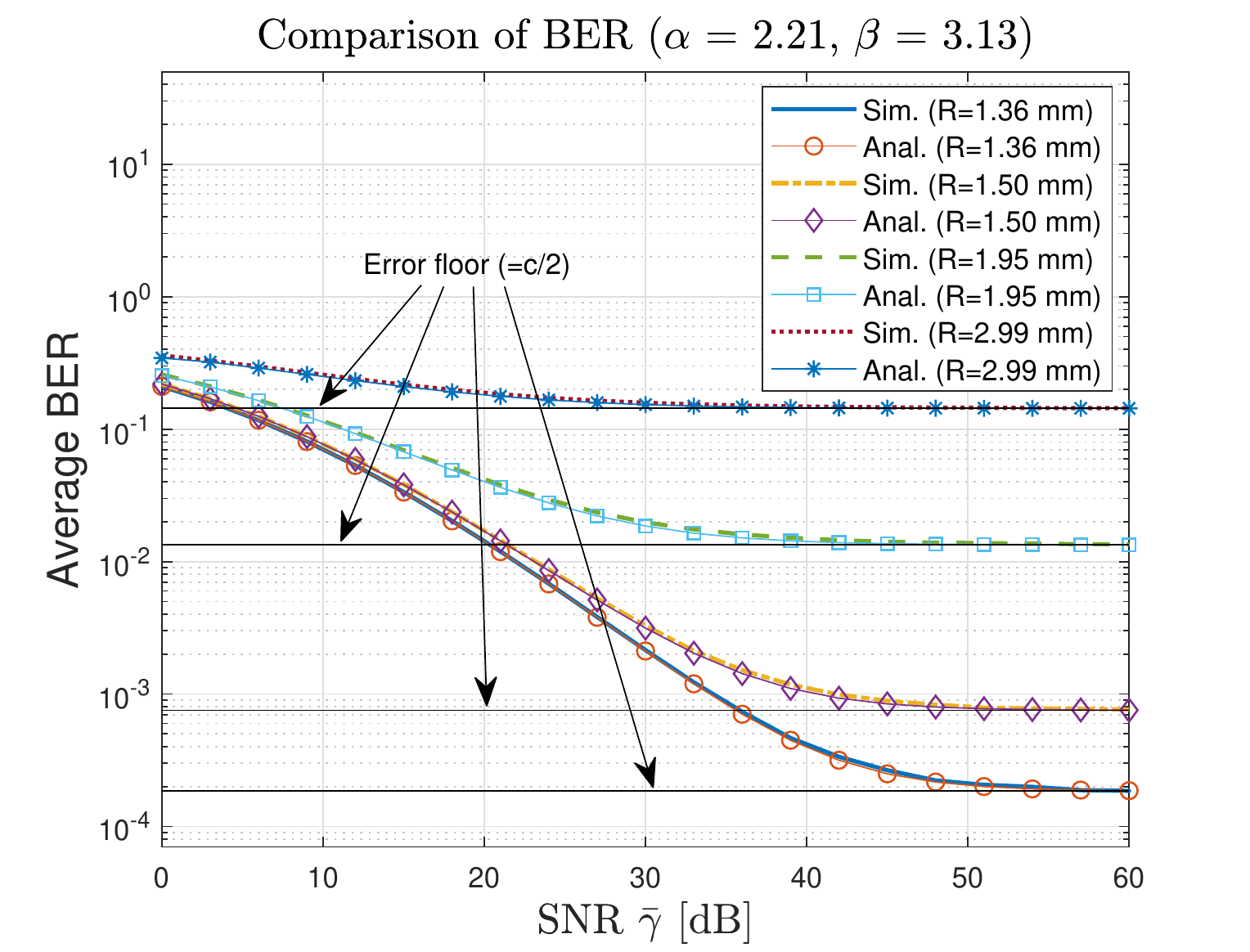}
	}
	\caption{The results of the performance analysis with respect to different horizontal movement under the given generation rate of $80$ [1/sec] and horizontal movement of $\sigma_x=5$ [mm] of a bubble.}
	\label{performance}
	\subfloat[Ergodic Capacity]{\includegraphics[width=3.5in]{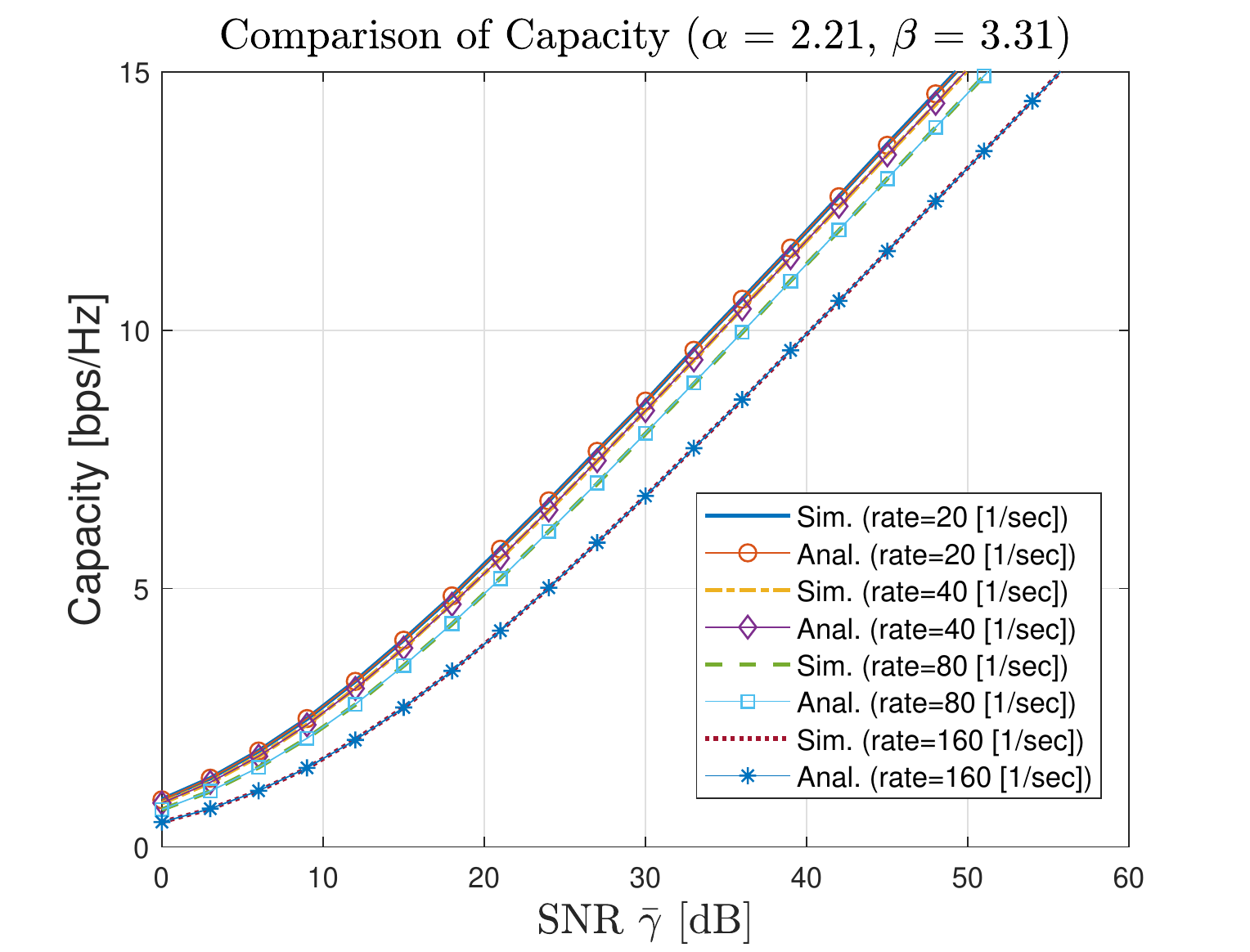}
	}
	\hfil
	\subfloat[Average BER]{\includegraphics[width=3.5in]{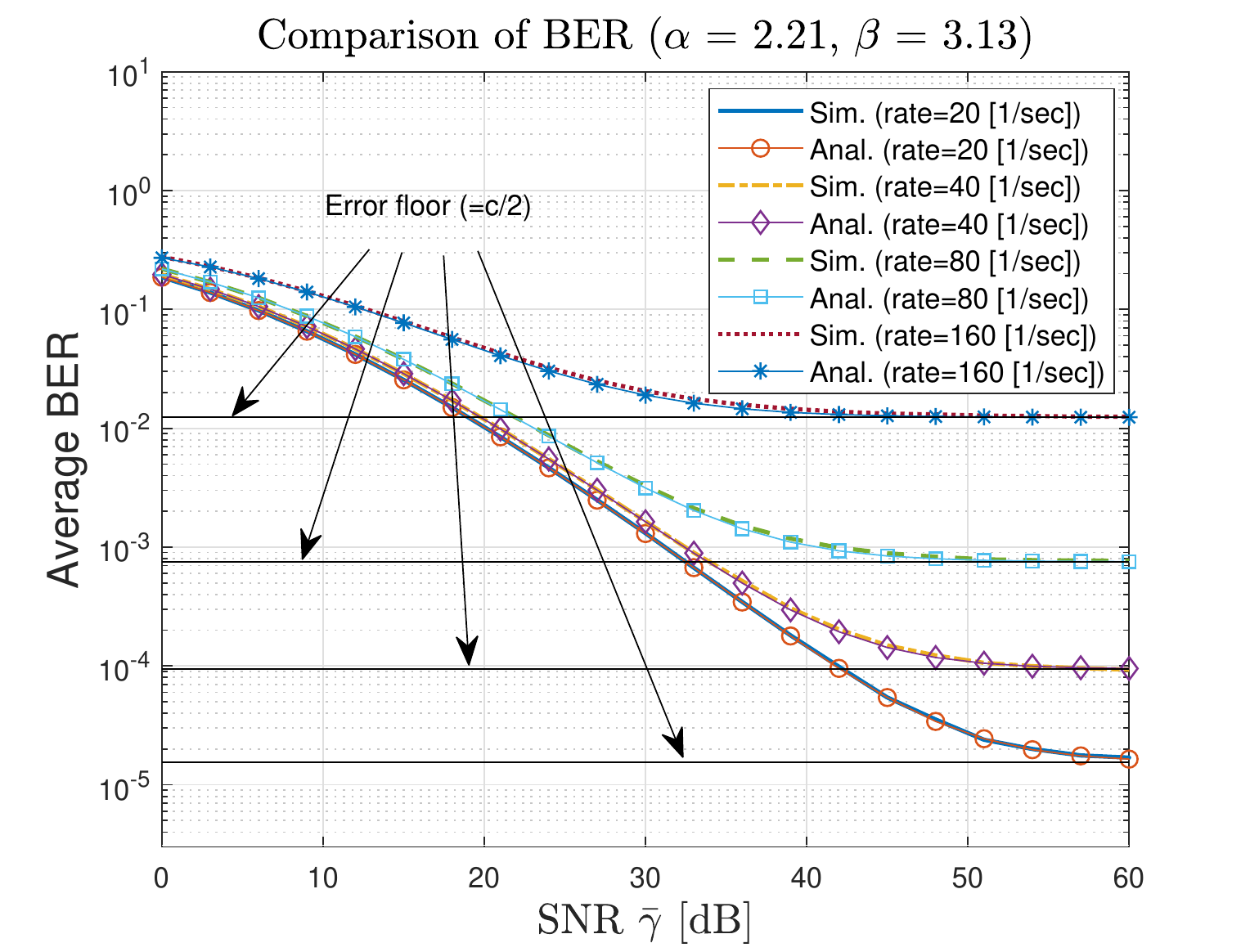}
	}
	\caption{The results of the performance analysis with respect to different bubble generation rate under the given mean radius of $\mu_R=1.50$ [mm] and horizontal movement of $\sigma_x=5$ [mm] of a bubble.}
	\label{performance_rate}
	%
	\centering
	\subfloat[Ergodic Capacity]{\includegraphics[width=3.5in]{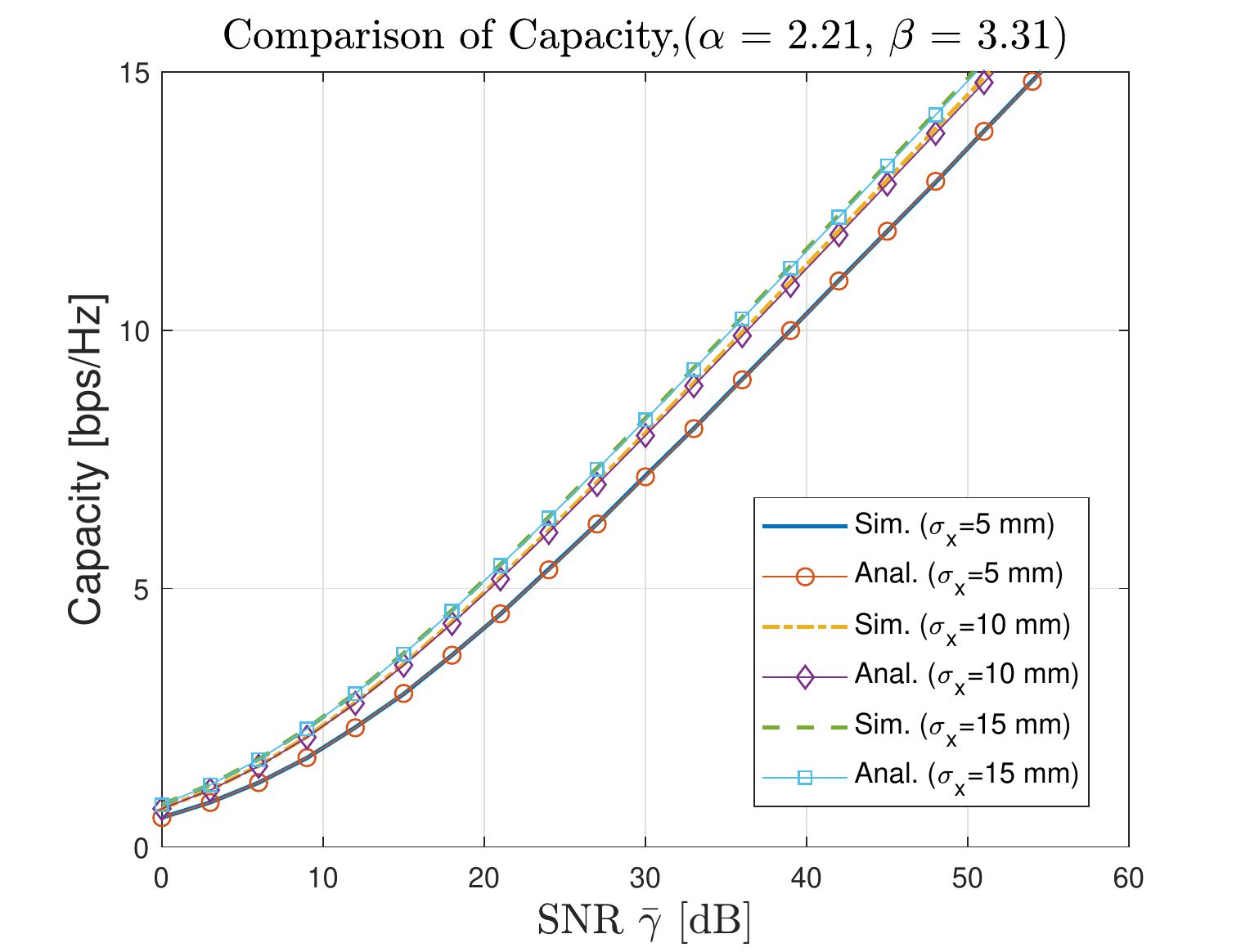}
	}
	\hfil
	\subfloat[Average BER]{\includegraphics[width=3.5in]{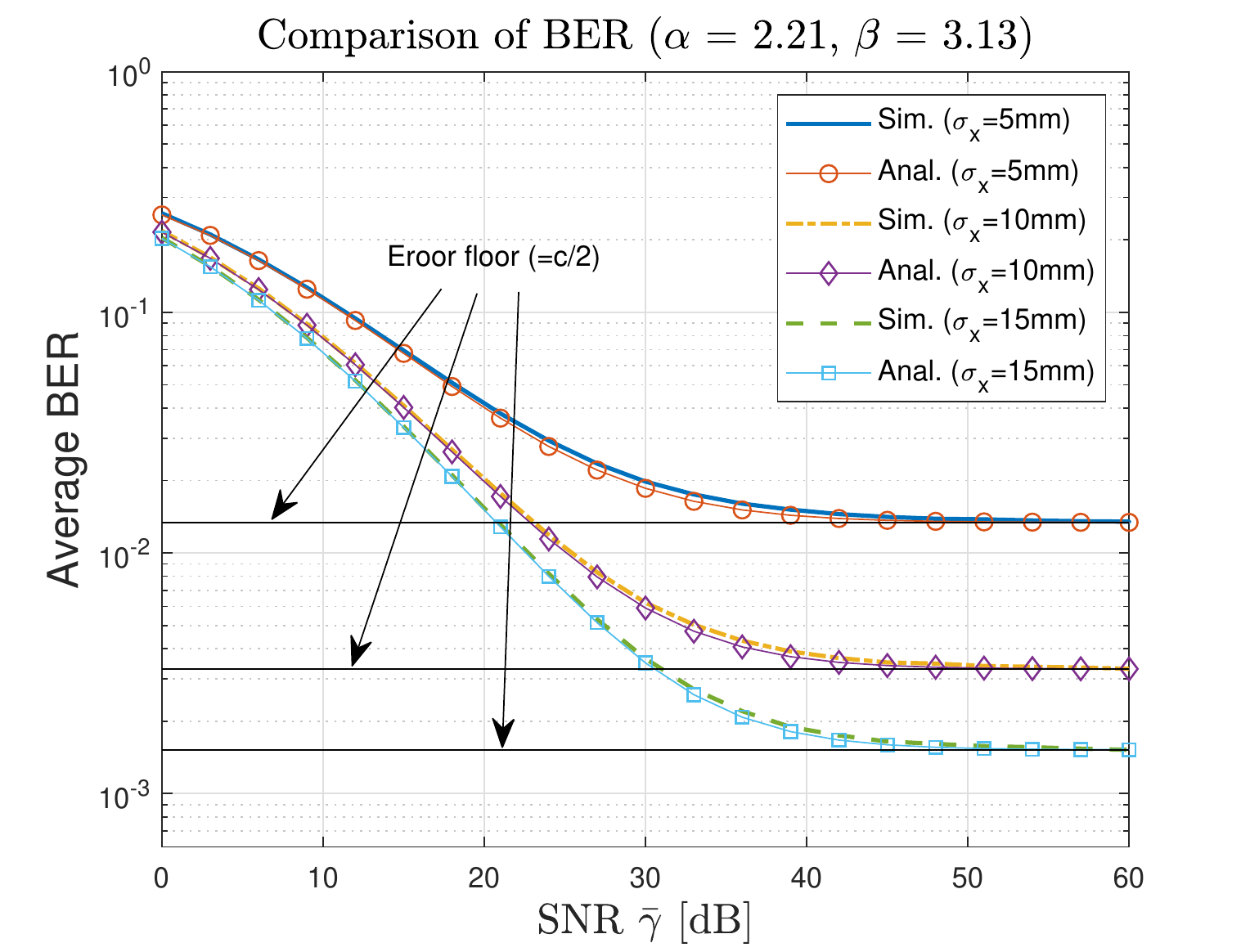}
	}
	\caption{The results of the performance analysis with respect to different horizontal movement under the given mean radius of $\mu_R=1.95$ [mm] and generation rate of $80$ [1/sec] of a bubble.}
	\label{performance_dev}
	
\end{figure*} 

\clearpage
\newpage

\subsection{Cases of the Obstructed Power}

\begin{figure}[t!]
	\centering
	\includegraphics[width=1\linewidth]{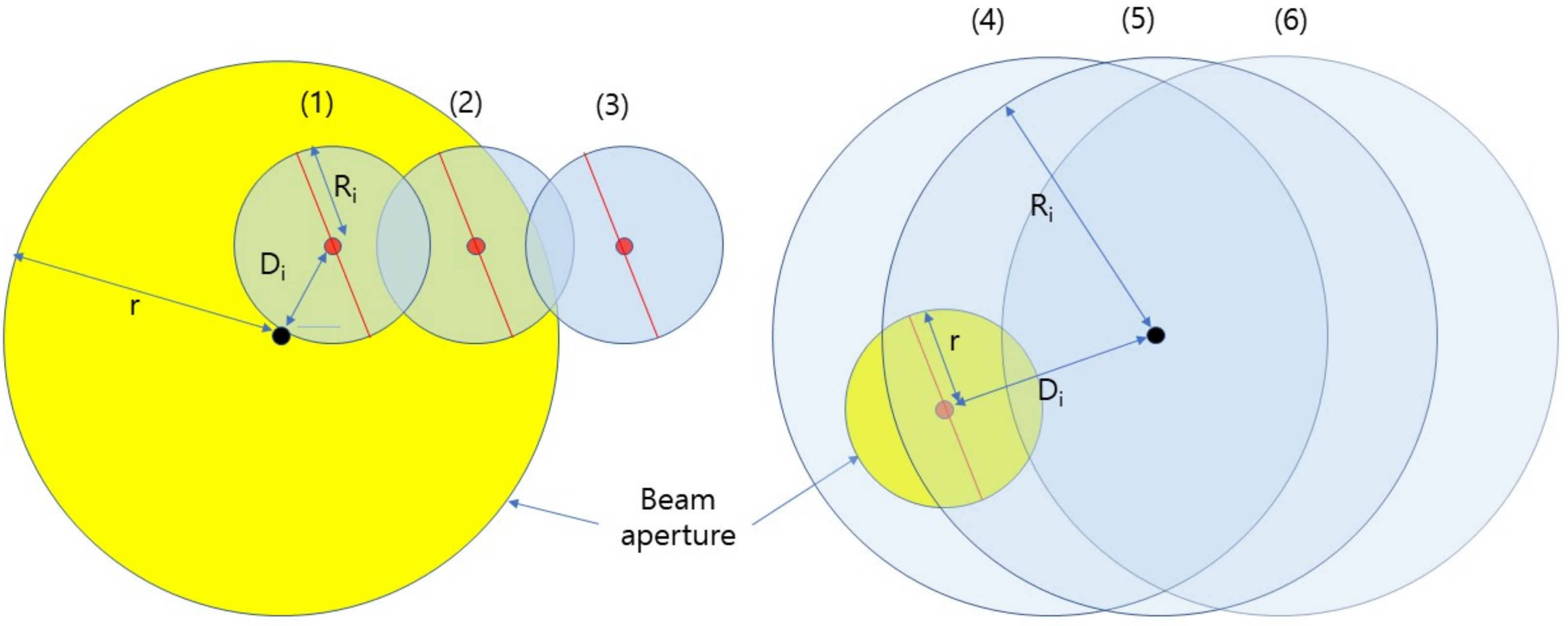}
	\caption{The six cases of bubbles overlapping the beam aperture.}
	\label{fig:bubblept}
\end{figure}

We can divide the obstructed power into six cases depending on where the bubble's location and size with respect to the beam. Fig.\ref{fig:bubblept} presents a pictorial description of each case. We will derive the six cases separately as follows.  

\begin{itemize}
	\item[(1)] $r \ge R_i, \, D_i \le r - R_i$: When the bubble is completely inside the beam area, 
	\begin{align}
	B_i^{(1)} & = b_i^{(1)}(X_i, R_i, T_i) \nonumber\\ 
	& = \int^{R_i}_{-R_i}\int^{\sqrt{R_i^2 - w^2} - D_i}_{- \sqrt{R_i^2 - w^2} - D_i}h(w,z)\,dz\,dw.
	\end{align}
	\item[(2)] $r \ge R_i, \, D_i > r - R_i, \, D_i^2 \le r^2 - R_i^2$: When the bubble partially overlaps with the beam aperture and one of the diameters passing through the center of bubble is entirely contained in the aperture,
	\begin{align}
	B_i^{(2)} & = b_i^{(2)}(X_i, R_i, T_i) \nonumber \\& = \int^{R_i}_{-R_i}\int^{\sqrt{R_i^2 - w^2} - D_i}_{- \sqrt{R_i^2 - w^2}  - D_i}h(w,z)\,dz\,dw \nonumber \\ 
	& -\int^{\sqrt{r^2 - \frac{R_i^2 - r^2 - D_i^2}{2D_i}}}_{-\sqrt{r^2 - \frac{R_i^2 - r^2 - D_i^2}{2D_i}}}\int^{- \sqrt{r^2 - w^2}}_{ -\sqrt{R_i^2 - w^2} - D_i}h(w,z)\,dz\,dw.
	\end{align}
	\item[(3)] $r \ge R_i, \, D_i^2 > r^2 - R_i^2, \, D_i \le r + R_i$: When the bubble partially overlaps with the beam aperture and no diameter of the bubble is contained in the aperture,
	\begin{align}
	B_i^{(3)}&=b_i^{(3)}(X_i, R_i, T_i) \nonumber\\
	&= \int^{\sqrt{r^2 - \frac{R_i^2 - r^2 - D_i^2}{2D_i}}}_{-\sqrt{r^2 - \frac{R_i^2 - r^2 - D_i^2}{2D_i}}}\int^{\sqrt{R_i^2 - w^2} - D_i}_{ - \sqrt{r^2 - w^2}}h(w,z)\,dz\,dw.
	\end{align}

	\item[(4)] $r < R_i, \, D_i \le R_i - r$  : When the beam aperture is completely inside the bubble area,
	\begin{eqnarray}
	B_i^{(4)} \!\!&\!\!=\!\!&\!\! b_i^{(4)}(X_i, R_i, T_i)\nonumber\\
	\!\!&\!\!=\!\!&\!\! \int^{r}_{-r}\int^{\sqrt{r^2 - w^2} }_{- \sqrt{r^2 - w^2} }h(w,z)\,dz\,dw \triangleq m.
	\end{eqnarray}
	\item[(5)] $r < R_i, \, D_i > R_i - r, \, D_i^2 \le R_i^2 - r^2$: When the bubble partially overlaps with the beam aperture and one of the diameters of the beam aperture is entirely contained in the bubble,
	\begin{align}
	B_i^{(5)} &= b_i^{(5)}(X_i, R_i, T_i) \nonumber\\ &= \int^{r}_{-r}\int^{\sqrt{r^2 - w^2} }_{- \sqrt{r^2 - w^2} }h(w,z)\,dz\,dw \nonumber\\ & - \int^{-\sqrt{r^2 - \frac{R_i^2 - r^2 - D_i^2}{2D_i}}}_{\sqrt{r^2 - \frac{R_i^2 - r^2 - D_i^2}{2D_i}}}\!\!\int^{ \sqrt{r^2 - w^2}}_{ \sqrt{R_i^2 - w^2} - D_i}h(w,z)\,dz\,dw.
	\end{align}
	\item[(6)] $r < R_i, \, D_i^2 > R_i^2 - r^2 , \, D_i \le r + R_i$: when the bubble overlaps with the beam aperture and no diameter of the beam aperture is contained in the bubble,
	\begin{align}
	B_i^{(6)} &= b_i^{(6)}(X_i, R_i, T_i) \nonumber \\
	&= \int^{\sqrt{r^2 - \frac{R_i^2 - r^2 - D_i^2}{2D_i}}}_{-\sqrt{r^2 - \frac{R_i^2 - r^2 - D_i^2}{2D_i}}}\int^{\sqrt{r^2 - w^2} - D_i}_{ - \sqrt{r^2 - w^2}}h(w,z)\,dz\,dw.
	\end{align}
	
\end{itemize}

\subsection{Calculations of Moments for Totally Obstructed Power $B$}
\begin{itemize}
	\item[(1)] The first moment of $B$
	
	The expectation of the $i$th bubble
	is given on the top of next page by \eqref{expectation}. 
	\begin{figure*}
	\begin{equation}
	\mathbb{E}[B_i] = \int^{iL}_{(i-1)L} \left[\int^{0.01}_{0}\left[\int^{\infty}_{- \infty}b_i(x_i,r_i,t_i)f_{X_i}(x_i)\,dx_i\right]\\
	f_{R_i}(r_i)\,dr_i\right]f_{T_i}(t_i)\,dt_i. \label{expectation}
	\end{equation}
	\hrule
	\end{figure*}
	We note that $f_{X_i}(x)$, $f_{R_i}(r)$, and $f_{T_i}(t)$ are the distributions of horizontal movement $X_i$, radius $R_i$, and generation time $T_i$, of air bubble, respectively, which are given in Section II. The expectation of the sum of the obstructed power by the $10/L$ bubbles, which are all the generated bubbles during the overall time duration of 10 [sec] at a generation rate $1/L$, is given by
	\begin{equation}
	\mathbb{E}[B] = \sum _{i = 1 }^{10 /L} \mathbb{E}[B_i]. \label{E_B} 
	\end{equation}

	\item[(2)] The second moment of $B$
	
	The second moment the $i$th bubble is given on the top of next page by \eqref{smoment}. 
	\begin{figure*}
	\begin{equation}
	\mathbb{E}[{B_i}^2] = \int^{iL}_{(i-1)L} \left[\int^{0.01}_{0}\left[\int^{\infty}_{- \infty}{b_i}^2(x_i,r_i,t_i)f_{X_i}(x_i)\,dx_i\right]\\f_{R_i}(r_i)\,dr_i\right]f_{T_i}(t_i)\,dt_i. \label{smoment}
	\end{equation}
	\hrule
	\end{figure*}
	The second moment of the sum of the obstructed power is represented as
	\begin{equation}
	\mathbb{E}[B^2] =\sum _{i = 1 }^{10/L}\mathbb{E}\left[{{B_i}^2}\right] +2 \sum _{i = 1 }^{10 /L}\sum _{j = 1 }^{10 /L}\mathbb{E}[{B}_i]\mathbb{E}[{B}_j]. \label{V_B}
	\end{equation}
	
\end{itemize}

\subsection{Parameter Estimation}
To begin, we assume the distribution of the power obstructed by air bubbles follows the form of $ a\delta(x)+ b f(x)$ $(x \ge 0)$, where $f(x)$ can be determined by any suitable density function. Here, $a$ is the probability of no power obstruction, i.e., no bubble obstructs the power. The probability of each bubble not obstructing the power can be calculated as
\begin{equation}
\int^{iL}_{(i-1)L} \int^{0.01}_{0}\int^{\infty}_{- \infty} \mathbbm{1}_{( \, r + {R_i},\,\infty)}({D_i})\,dxf_{R_i}(r_i)\,dr_if_{T_i}(t)dt_i
\end{equation}
Therefore, $a$ is given on the top of next page by \eqref{a}.
\begin{figure*}
\begin{equation}
a = \prod_{i = 1 }^{10/L} \int^{iL}_{(i-1)L} \left[\int^{0.01}_{0}\left[\int^{\infty}_{- \infty} \mathbbm{1}_{( \, r + {R_i},\,\infty)}({D_i})\,dx\right]f_{R_i}(r_i)\,dr_i\right]f_{T_i}(t_i)\,dt_i. \label{a}
\end{equation}
\hrule
\end{figure*}
We note that $b = 1- a$, because $\int_0^{\infty} a\delta(x)+ bf(x) \,dx = 1$.

Based on the observation of right tail distribution from simulation data, we now model $f(x)$ as the Weibull distribution, which is represented as
\setcounter{eqcnt}{\value{equation}}
\setcounter{equation}{12}
\begin{equation}
f(x) = f_W(x) = \left\{
\begin{array}{ll}
\displaystyle \frac{k}{\lambda}\left(\frac{x}{\lambda}\right)^{k-1} e^{(x/ \lambda)^k}, & x \ge 0, \\
0, & x < 0.
\end{array}
\right.
\end{equation}

Here, we have two parameters $k$ and $\lambda$ to estimate. The expectation and the second moment of the distribution $ a\delta(x)+ b f_W(x)$ are $b\lambda\Gamma\left(1+\frac{1}{k}\right)$ and $b\lambda^2\Gamma\left(1+\frac{2}{k}\right)$, respectively. Letting $\mathbb{E}[B] = b\lambda\Gamma\left(1+\frac{1}{k}\right)$ and $\mathbb{E}[B^2] =b\lambda^2\Gamma\left(1+\frac{2}{k}\right)$, we obtain $k$ by solving the following equation.
\begin{equation}
\mathbb{E}^2[B]\Gamma\left(1+\frac{2}{k}\right) = b\mathbb{E}[B^2]\Gamma^2\left(1+\frac{1}{k}\right).
\end{equation}
Then, $\lambda$ can be obtained by solving one of two moments with the obtained $k$.

\bibliographystyle{IEEEtran} 
\bibliography{References}

\begin{thebibliography}{10}
\providecommand{\url}[1]{#1}
\csname url@samestyle\endcsname
\providecommand{\newblock}{\relax}
\providecommand{\bibinfo}[2]{#2}
\providecommand{\BIBentrySTDinterwordspacing}{\spaceskip=0pt\relax}
\providecommand{\BIBentryALTinterwordstretchfactor}{4}
\providecommand{\BIBentryALTinterwordspacing}{\spaceskip=\fontdimen2\font plus
\BIBentryALTinterwordstretchfactor\fontdimen3\font minus
  \fontdimen4\font\relax}
\providecommand{\BIBforeignlanguage}[2]{{%
\expandafter\ifx\csname l@#1\endcsname\relax
\typeout{** WARNING: IEEEtran.bst: No hyphenation pattern has been}%
\typeout{** loaded for the language `#1'. Using the pattern for}%
\typeout{** the default language instead.}%
\else
\language=\csname l@#1\endcsname
\fi
#2}}
\providecommand{\BIBdecl}{\relax}
\BIBdecl

\bibitem{key6}
H.~{Kaushal} and G.~{Kaddoum}, ``Underwater optical wireless communication,''
  \emph{IEEE Access}, vol.~4, pp. 1518--1547, Apr. 2016.

\bibitem{Shen16}
C.~Shen, Y.~Guo, H.~M. Oubei, T.~K. Ng, G.~Liu, K.-H. Park, K.-T. Ho, M.-S.
  Alouini, and B.~S. Ooi, ``20-meter underwater wireless optical communication
  link with 1.5 {Gbps} data rate,'' \emph{Opt. Express}, vol.~24, no.~22, pp.
  25\,502--25\,509, Oct. 2016.

\bibitem{key8}
C.~Li, K.-H. Park, and M.-S. Alouini, ``On the use of a direct radiative
  transfer equation solver for path loss calculation in underwater optical
  wireless channels,'' \emph{IEEE Wireless Commun. Lett.}, vol.~4, no.~5, pp.
  561--564, Oct. 2015.

\bibitem{key2}
H.~Gao and H.~Zhao, ``A fast-forward solver of radiative transfer equation,''
  \emph{Transport Theory and Statistical Physics}, vol.~38, no.~3, pp.
  149--192, 2009.

\bibitem{key3ex}
E.~Illi, F.~E. Bouanani, K.-H. Park, F.~Ayoub, and M.-S. Alouini, ``An improved
  accurate solver for the time-dependent rte in underwater optical wireless
  communications,'' \emph{IEEE Access}, vol.~7, pp. 96\,478--96\,494, Jul.
  2019.

\bibitem{key9}
R.~Leathers, T.~Downes, C.~O.~Davis, and C.~Mobley, ``Monte {C}arlo radiative
  transfer simulations for ocean optics: A practical guide,'' \emph{Naval
  Research Laboratory}, p.~54, Sep. 2004.

\bibitem{key10}
Z.~{Zeng}, S.~{Fu}, H.~{Zhang}, Y.~{Dong}, and J.~{Cheng}, ``A survey of
  underwater optical wireless communications,'' \emph{IEEE Commun. Surveys
  Tuts.}, vol.~19, no.~1, pp. 204--238, Firstquarter 2017.

\bibitem{bubbles}
\BIBentryALTinterwordspacing
{Jeffrey B. Graham Perspectives on Ocean Science Lecture Series}. (2018)
  There's more to ocean bubbles than you might think. [Online]. Available:
  \url{https://www.uctv.tv/shows/Theres-More-to-Ocean-Bubbles-Than-You-Might-Think-33487}
\BIBentrySTDinterwordspacing

\bibitem{key11}
E.~{Zedini}, H.~M. {Oubei}, A.~{Kammoun}, M.~{Hamdi}, B.~S. {Ooi}, and M.-S.
  {Alouini}, ``Unified statistical channel model for turbulence-induced fading
  in underwater wireless optical communication systems,'' \emph{IEEE Trans.
  Commun.}, vol.~67, no.~4, pp. 2893--2907, Apr. 2019.

\bibitem{key23}
M.~V. Jamali, A.~Mirani, A.~Parsay, B.~Abolhassani, P.~Nabavi, A.~Chizari,
  P.~Khorramshahi, S.~Abdollahramezani, and J.~A. Salehi, ``Statistical studies
  of fading in underwater wireless optical channels in the presence of air
  bubble, temperature, and salinity random variations,'' \emph{IEEE Trans.
  Commun.}, vol.~66, no.~10, pp. 4706--4723, Oct. 2018.

\bibitem{key12}
H.~M. Oubei, R.~T. ElAfandy, K.-H. Park, T.~K. Ng, M.-S. Alouini, and B.~S.Ooi,
  ``Performance evaluation of underwater wireless optical communications links
  in the presence of different air bubble populations,'' \emph{IEEE Photonics
  J.}, vol.~9, no.~2, pp. 1--9, Apr. 2017.

\bibitem{key13}
S.~H. Park, C.~Park, J.~Lee, and B.~Lee, ``A simple parameterization for the
  rising velocity of bubbles in a liquid pool,'' \emph{Nuclear Engineering and
  Technology}, vol.~49, no.~4, pp. 692--699, Oct. 2017.

\bibitem{key22}
K.~O. Bowman and L.~R. Shenton, ``Method of moments,'' \emph{Encyclopedia of
  statistical sciences}, vol.~5, pp. 2092--2098, 1985.

\bibitem{key18}
A.~A. Farid and S.~Hranilovic, ``Outage capacity optimization for free-space
  optical links with pointing errors,'' \emph{J. Lightwave Technol.}, vol.~25,
  no.~7, pp. 1702--1710, July 2007.

\bibitem{key19}
I.~S. {Ansari}, F.~{Yilmaz}, and M.-S. {Alouini}, ``Performance analysis of
  free-space optical links over málaga ($\mathcal{M} $) turbulence channels
  with pointing errors,'' \emph{IEEE Trans. Wireless Commun.}, vol.~15, no.~1,
  pp. 91--102, Jan. 2016.

\bibitem{key21}
V.~S. Adamchik and O.~I. Marichev, ``The algorithm for calculating integrals of
  hypergeometric type functions and its realization in reduce system,'' in
  \emph{Proc. International Symposium on Symbolic and Algebraic Computation
  (ISSAC' 90)}, 1990, pp. 212--224.

\bibitem{key20}
H.~G. Sandalidis, T.~A. Tsiftsis, G.~K. Karagiannidis, and M.~Uysal, ``{BER}
  performance of {FSO} links over strong atmospheric turbulence channels with
  pointing errors,'' \emph{IEEE Commun. Lett.}, vol.~12, no.~1, pp. 44--46,
  Jan. 2008.

\bibitem{key15}
M.~Elamassie and M.~Uysal, ``Vertical underwater {VLC} links over cascaded
  {G}amma-{G}amma turbulence channels with pointing errors,'' in \emph{Proc.
  IEEE International Black Sea Conf. on Commun. Netw. (IEEE BlackSeaCom 2019)},
  Sochi, Russia, Jun. 2019.

\bibitem{key16}
L.~C. Andrews and R.~L. Phillips, \emph{Laser Beam Propagation through Random
  Media}, 2nd~ed.\hskip 1em plus 0.5em minus 0.4em\relax Bellingham, WA, USA:
  SPIE, 2005.

\bibitem{key17}
M.~Cheng, L.~Guo, and Y.~Zhang, ``Scintillation and aperture averaging for
  {G}aussian beams through non-{K}olmogorov maritime atmospheric turbulence
  channels,'' \emph{Opt. Express}, vol.~23, no.~25, pp. 32\,606--32\,621, Dec.
  2015.

\end{thebibliography}

\end{document}